\documentclass[reprint,prx,aps,superscriptaddress]{revtex4-2}
\usepackage{times}
\usepackage{graphicx}
\usepackage{amsmath, braket, amsfonts}
\usepackage{amssymb,tabularx}
\usepackage{sidecap}
\usepackage{cancel}
\usepackage{bm, color, ulem}
\usepackage{ragged2e}
\usepackage{svg}
\usepackage{wrapfig,lipsum,booktabs}

\usepackage{siunitx}

\newcommand{\be}{\begin{equation}}
\newcommand{\ee}{\end{equation}}

\newcommand{\Tr}{\operatorname{Tr}}
\newcommand{\bbI}{\mathbb{I}}
\newcommand{\calL}{\mathcal{L}}
\newcommand{\calO}{\mathcal{O}}

\newcommand{\q}{\boldsymbol{q}}

\definecolor{myforestgreen}{RGB}{34,139,34}

\DeclareUnicodeCharacter{03BD}{{$\nu$}}

\makeatletter
\def\maketitle{
\@author@finish
\title@column\titleblock@produce
\suppressfloats[t]}
\makeatother

\begin{document}
\title{Entropy of strongly correlated electrons in a partially filled Landau level}
\author{Alexandre Assouline}
\email{These authors contributed equally}
\affiliation{Department of Physics, University of California at Santa Barbara, Santa Barbara CA 93106, USA}
\affiliation{CNRS, Institut Néel, 38000 Grenoble, France}
\author{Taige Wang}
\email{These authors contributed equally}
\affiliation{Department of Physics, University of California, Berkeley, California 94720, USA}
\affiliation{Material Science Division, Lawrence Berkeley National Laboratory, Berkeley, California 94720, USA}
\author{Heun Mo Yoo}
\email{These authors contributed equally}
\affiliation{Department of Physics, University of California at Santa Barbara, Santa Barbara CA 93106, USA}
\author{Ruihua Fan}
\affiliation{Department of Physics, University of California, Berkeley, California 94720, USA}
\author{Fangyuan Yang}
\affiliation{Department of Physics, University of California at Santa Barbara, Santa Barbara CA 93106, USA}
\author{Ruining Zhang}
\affiliation{Department of Physics, University of California at Santa Barbara, Santa Barbara CA 93106, USA}
 \author{Takashi Taniguchi}
 \affiliation{International Center for Materials Nanoarchitectonics,
 National Institute for Materials Science,  1-1 Namiki, Tsukuba 305-0044, Japan}
 \author{Kenji Watanabe}
 \affiliation{Research Center for Functional Materials,
 National Institute for Materials Science, 1-1 Namiki, Tsukuba 305-0044, Japan}
 \author{Michael P. Zaletel}
\affiliation{Department of Physics, University of California, Berkeley, California 94720, USA}
\affiliation{Material Science Division, Lawrence Berkeley National Laboratory, Berkeley, California 94720, USA}
\author{Andrea F. Young}
\email{andrea@physics.ucsb.edu}
 \affiliation{Department of Physics, University of California at Santa Barbara, Santa Barbara CA 93106, USA}

\date{\today}

\begin{abstract}

We use high-resolution chemical potential measurements to extract the entropy of monolayer and bilayer graphene in the quantum Hall regime via the Maxwell relation $\left.\frac{d\mu}{dT}\right|_N = -\left.\frac{dS}{dN}\right|_T$. Measuring the entropy from $T=\SI{300}{\kelvin}$ down to $T=\SI{200}{\milli\kelvin}$, we identify the sequential emergence of quantum Hall ferromagnetism, fractional quantum Hall states (FQH), and various charge orders by comparing the measured entropy in different temperature regimes with theoretical models. At the lowest temperature of $T \approx \SI{200}{\milli\kelvin}$ we perform a detailed study of the entropy near even-denominator fractional quantum Hall states in bilayer graphene, and comment on the possible topological origin of the observed excess entropy.
\end{abstract}

\maketitle

\section{Introduction}
{In a correlated electron system, the temperature-dependent entropy gives direct experimental access to the degrees of freedom the system can explore at a given energy scale.
At high temperatures, the entropy is typically dominated by the motional degrees of freedom of individual particles. However, at lower temperature, the entropy is determined by the long-wavelength collective excitations.
Entropy thus provides a unique fingerprint of the energetic hierarchy of a many-body system.
Entropy measurements are of particular interest in flat electronic bands where strong interactions can give rise to a rich variety of electronic orders hosting a correspondingly rich array of excitations.
For example, in moire- and rhombohedral graphene systems, entropy measurements have been used to probe the spin- and valley excitations, revealing a regime of strongly fluctuating magnetic moments at finite temperature~\cite{saito_isospin_2021,rozen_entropic_2021,liu_isospin_2022,holleis_fluctuating_2025}.

Landau levels offer an even more ideal flat band system in which entropy measurements can directly probe the nature of the emergent excitations.
For example, entropy measurements are predicted to provide a signature of the topological ground state degeneracy expected in the vicinity of nonabelian fractional quantum Hall states as are thought to occur in a half-filled first orbital Landau level. Specifically, it was proposed~\cite{cooper_observable_2009} that for electron density tuned near such fillings---where the ground state is expected to be a Wigner crystal of nonabelian anyons with charge $e/4$---the nonabelian statistics lead to an additional entropy of $\ln(\sqrt{2})$ per fractionally charged quasiparticle. This entropy arises from the state space of the quantum bits delocalized between each pair of quasiparticles.
When the motional entropy of the anyons is quenched, the entropy is expected to saturate at $S_{\mathrm{topo}} \approx k_B N_q\ln(\sqrt{2})$, where $N_q$ is the number of non-Abelian quasiparticles. Equating $\mathrm{d}N_q / \mathrm{d}N = 4$ (where $N$ is the number of electrons), this should result in a low-$T$ plateau of  $\mathrm{d}S/\mathrm{d}N =  4 k_B \ln\sqrt{2}$, revealing the topological degeneracy native to such states.

Previously, several experiments have reported entropy measurements in partially filled Landau levels (LL), including in silicon~\cite{kuntsevich_strongly_2015} and GaAs quantum wells~\cite{chickering_thermopower_2010,chickering_thermoelectric_2013,schmidt_specific_2017}. However, these experiments were typically restricted to a narrow range of filling factors and temperature. In the current work, we probe the entropy of quantum Hall states in mono- and bilayer graphene over four orders of magnitude of temperature, and across a range of density spanning several LLs.
To determine the entropy, we first measure the temperature dependent chemical potential $\mu(N, T)$ using a proximal charge sensor~\cite{eisenstein_negative_1992,eisenstein_compressibility_1994,yang_experimental_2021,park_flavour_2021,saito_isospin_2021,rozen_entropic_2021,yang_cascade_2023,assouline_energy_2024}. As shown in the inset of Fig.~\ref{fig:1}a, our experimental setup uses a multilayer heterostructure that includes a sample layer consisting of either mono- or bilayer graphene with applied voltage $v_s$, and a sensor layer consisting of a grounded graphene monolayer in which Corbino transport is measured. Adjusting the top and bottom gate voltages ($v_t$ and $v_b$, respectively) to keep the density in the sensor layer fixed at a fractional quantum Hall conductance minimum allows the chemical potential $\mu$ of the sample layer to be measured as a function of Landau level filling $\nu$ and $T$, as described previously~\cite{yang_experimental_2021}.
The entropy $S$ then follows from the Maxwell relation,
\begin{equation} \label{eq:Maxwell}
    \frac{\mathrm{d}\mu}{\mathrm{d}T}\bigg|_N = -\frac{\mathrm{d}S}{\mathrm{d}N}\bigg|_T,
\end{equation}
where in the experiment  is approximated by the finite difference between $\mu$ curves taken at different
\begin{figure*}[ht]
    \centering
    \includegraphics[width = \textwidth]{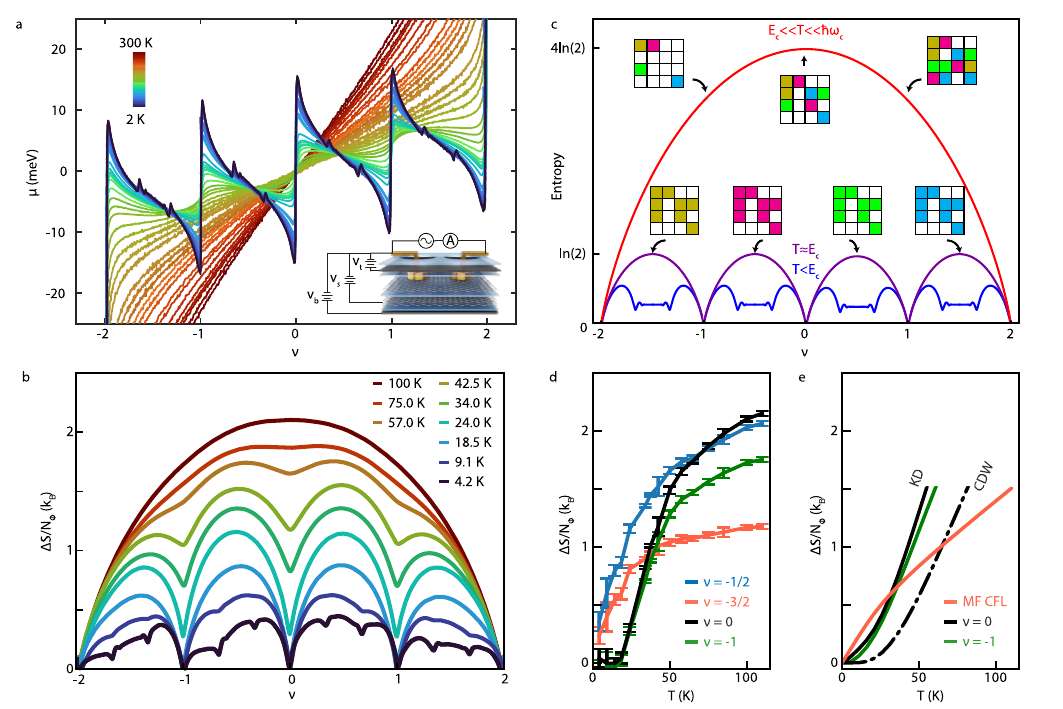}
    \caption{\textbf{Entropy in a partially filled monolayer graphene Landau level.}
    \textbf{(a)} The chemical potential $\mu$ of monolayer graphene at $B = \SI{11.5}{\tesla}$ at temperatures ranging from \SI{2}{K} to \SI{300}{K}.
    \textbf{(b)} The entropy per flux determined from the temperature dependence of $\mu(\nu)$ in panel a.
    \textbf{(c)} Schematic of the entropy per flux in the high (red), intermediate (purple) and low (blue) temperature regimes. The domes are associated with the occupational entropy of four isospin flavors ($T \gg E_C$) or a single isospin flavor ($T \sim E_C$). Available motional states are depicted on a grid, with the colors representing isospin flavors.
    \textbf{(d)} Measured T-dependent entropy at integer and half-integer fillings.
    \textbf{(e)}
    Temperature dependence of the entropy from the non-interacting magnon model at integer fillings, and from a mean-field composite Fermi liquid theory at half-integer filling. The dash-dotted line assumes $SU(4)$ symmetric Coulomb interactions, while the solid line accounts for  realistic isospin anisotropy (see supplementary information).}
    \label{fig:1}
\end{figure*}

\section{High-temperature positional entropy}

Fig.~\ref{fig:1}a presents  $\mu$ for monolayer graphene at $B = \SI{11.5}{\tesla}$ across temperatures ranging from \SI{2.7}{K} to \SI{300}{K}.  At low temperatures, $\mu$ shows numerous well-understood signatures of electronic correlations including sudden jumps in $\mu$ at Landau level fillings factor $\nu = -2, -1, 0, 1, 2$ associated with quantum Hall ferromagnetic gaps and negative electronic compressibility ($\partial \mu/\partial \nu < 0$)~\cite{eisenstein_compressibility_1994}. To extract the entropy, we  approximate the derivative in Eq. \eqref{eq:Maxwell} by the difference between $\mu$  curves obtained at slightly different temperatures, $\mathrm{d}S/\mathrm{d}N=-d\mu/dT\approx -\delta \mu/\delta T$, where $\delta \mu = \rm \mu_{hot}-\mu_{cold}$ and $\delta T= T_{\text{hot}} - T_{\text{cold}}$.
The entropy we measure is thus approximately the average entropy across this temperature window.
We then obtain the entropy difference between two filling factors via integration, $\Delta S / N_{\phi}= \int_{\nu_1}^{\nu_2} (\mathrm{d}S/\mathrm{d}N ) d\nu$. Here $N_\phi$ is the number of flux quanta piercing the sample.
This procedure involves two arbitrary constants that must be constrained experimentally.
First, a reference filling must be chosen such that $\Delta S$ provides a good representation of the total entropy $S$.
Second, strictly speaking we measure the difference in chemical potential between the sample and the ``sensor'' layer, $\mu(\nu, T) - \mu_{\text{sens}}(T)$.
Since the sensor layer is kept at fixed $\nu$, $\mu(\nu, T)$ is known up to a density-independent but temperature-dependent constant, $\mu_{\text{sens}}(T)$.
Upon integration, this results in a linear background added to the entropy, $S / N_\phi\to  S / N_\phi - (\nu - \nu_1)  \partial \mu_{\text{sens}}(T) / \partial T$.
To remove this background, we require that the entropy is negligible for $T < \SI{100}{K}$  at  $\nu = \pm 2$. This is justified by theoretical calculations using the MLG Landau level spectrum, which give $S(\nu = \pm 2, T=\SI{100}{\kelvin}) / N_\phi = \SI{0.02}{k_B}$.
As a check, we note that the $S$ resulting from integration shown in Fig.~\ref{fig:1}b displays a  high degree of particle-hole symmetry about charge neutrality; this expected symmetry would be spoiled if the offset $\mu_{\text{sens}}(T)$ were chosen incorrectly.

At \SI{100}{K}, $S$ exhibits a single dome with maximum at $\nu=0$ as shown in Fig.~\ref{fig:1}b.
As $T$ is reduced, the entropy is strongly suppressed at integer fillings, so that by $T \approx \SI{20}{K}$ the measured $S$ shows four approximately identical domes centered at each half-integer $\nu$. At $T= \SI{4.2}{K}$, additional entropy minima develop at values of $\nu$ associated with the onset of fractional quantum Hall states.
These findings are consistent with the energy hierarchy expected within an interacting partially filled Landau level~\cite{dean_fractional_2020}, illustrated in Fig.~\ref{fig:1}c.
Our experiment operates at $T$ well below the cyclotron gap $\Delta_c \sim \SI{1500}{K}$, so that electronic states are restricted to the monolayer graphene zero energy Landau level (zLL).
When $T$ is comparable to the Coulomb interaction scale $E_C \sim \SI{500}{K}$, electrons can freely occupy the nearly-degenerate $4 N_{\phi}$ orbitals of the zLL, where $N_{\phi}$ represents the number of magnetic flux quanta and the factor of four accounts for the combined spin- and valley degeneracy of each orbital state. At these high temperatures, the $S(\nu)$ corresponds to the number of ways to distribute the $(\nu + 2) N_\phi$ electrons among the $4 N_\phi$ available states:
\begin{align}
S/N_\phi &= k_B \ln \Omega /N_\phi = k_B \ln \binom{4 N_\phi}{(\nu + 2) N_\phi} /N_\phi  \\
&= k_B  \left [ (2 + \nu) \ln(2+ \nu) + (2 - \nu) \ln(2 - \nu) \right]
\label{eq:binomial}
\end{align}
i.e.  a single ``dome'' across the entire zLL.
In this model, the entropy per flux quantum peaks at $S / N_\phi= 4\ln(2) k_B\approx 2.8k_B$ at $\nu=0$.
At $\SI{100}{K}$, the experimental data shows a value somewhat below this, $S / N_\phi \approx 2.1 k_B$. This reduction is to be expected, as $E_C$ is not entirely negligible in comparison to $T$.

As we lower the temperature, the entropy becomes sensitive to correlation effects which favor certain many-body states. Notably, even when isospin symmetry is preserved by the Hamiltonian---so that $E_C$ is the only energy scale relevant to the partially filled LL energetics--a separation of energy scales occurs between ferromagnetic correlations at integer filling and the formation of more subtle fractional quantum Hall phases. In a  simplified picture, for temperature somewhat below $E_C$, electrons near the Fermi level occupy only one isospin flavor, with other isospin flavors separated by an energy gap. This energy gap arises through a combination of exchange,  Zeeman energies, and $SU(4)$-breaking anisotropies of the interaction.
In this limit, the problem is reduced to filling $\tilde{\nu} N_\phi$ states out of $N_\phi$ within that flavor, where $\tilde{\nu} = \nu- \lfloor \nu\rfloor$, and the entropy is described by four distinct domes with peak entropy per flux quantum of $S/N_\phi = \ln(2) k_B$:
\begin{equation} \label{eq:1dome}
S/N_\phi = k_B  \left [ \tilde{\nu} \ln \tilde{\nu} + (1 - \tilde{\nu}) \ln(1 - \tilde{\nu}) \right]
\end{equation}
As the temperature is reduced by another order of magnitude, fractional quantum Hall insulators, Wigner crystals, or other correlated states begin to emerge at fractional fillings, further suppressing the entropy.

To allow a quantitative comparison to theoretical modeling, we focus the remainder of the manuscript on specific regimes of $\nu$ and $T$ of progressively lower energy scale, beginning with integer filling within the ZLL.

\section{Integer quantum Hall ferromagnetism}
Fig.~\ref{fig:1}d shows $\Delta S/N_\phi$ as a function of $T$ at $\nu=-1$ and $\nu=0$ as well as at $\nu=-1/2$ and $\nu=-3/2$.  At the integer fillings, the entropy falls below our noise limit at low temperatures, becoming undetectable below $T \approx \SI{10}{\kelvin}$.   Notably, the entropy at $\nu=0$ and $\nu=1$ shows quantitatively similar behavior at low temperatures. In contrast, the half-integer states exhibit significantly higher entropy at low temperatures, consistent with gapless composite Fermi liquid (CFL) phases, which we will return to below.

To understand the low-$T$ IQH entropy, we appeal to the theory of quantum Hall ferromagnetism, which predicts that electrons spontaneously break the approximate $SU(4)$ symmetry relating spins and valleys, resulting in low-energy magnon modes.
When the $SU(4)$ symmetry is exact, these modes are gapless, with a $k^2$ dispersion that results in linear-$T$ entropy.
However, in the experiment $SU(4)$ is broken by spin and valley Zeeman fields, as well as small $SU(4)$-anisotropies of the Coulomb interaction, all of which can gap out or modify the magnon dispersion.
When considering \emph{only} the Zeeman effects, the magnons~\cite{wei_electrical_2018, zhou_solids_2019, assouline_excitonic_2021, pierce_thermodynamics_2022, zhou_strong-magnetic-field_2021} will be gapped by the spin Zeeman energy ($\Delta_Z = \SI{15}{K}$ at 11.5T) and the $A-B$ sublattice splitting induced by the hBN substrate, which acts as a valley Zeeman field in the zLL~\cite{hunt_massive_2013} ($\Delta_{AB} = \SI{80}{K}$ for this device, which was measured previously in ~\cite{yang_experimental_2021}).
At $\nu = -1$, the state would thus spin and valley polarize, with minimal magnon gap  $\Delta_Z$, while at  $\nu = 0$, the state would be a valley-polarized, spin-singlet (referred to in the literature as the CDW state~\cite{kharitonov_canted_2012}) with a magnon gap equal to $\Delta_{AB} = \SI{80}{K}$. To make this comparison quantitative, Fig.~\ref{fig:1}e shows the entropy calculated in a low-temperature expansion based on the thermal population of  magnon modes.
To do so we first use the single-mode approximation to obtain the magnon dispersion $\epsilon_n(k)$, and then in the spirit of the Debye model we  compute the resulting free-boson entropy, keeping modes up to a momentum scale set by the inverse magnetic length.
As shown in Fig.\ref{fig:1}e, when accounting only for Zeeman effects, the entropy of a valley-polarized $\nu = 0$ ``CDW'' state  is suppressed relative to $\nu = -1$, in line with the above discussion and in contrast to experiment~\cite{kharitonov_phase_2012, atteia_su4_2021}.

The comparatively large entropy at $\nu = 0$ may be accounted for by considering the anisotropy of the Coulomb interaction, which is known to favor a  Kekul\'e or canted-antiferromagnetic order at charge neutrality, rather than the valley-polarized  state,  under similar experimental conditions~\cite{kharitonov_phase_2012,das_coexistence_2022,zibrov_even-denominator_2018,coissard_imaging_2022,liu_visualizing_2022,kharitonov_canted_2012}.
These phases spontaneously break U(1) valley and/or spin rotation symmetry, and thus support gapless magnon (i.e. Goldstone) modes even in the presence of a sublattice splitting.
As shown in Fig.~\ref{fig:1}e, where we assume a spin-singlet Kekul\'e order, accounting for the contribution of this gapless mode to the entropy significantly increases the entropy at $\nu=0$ to a similar level as at $\nu=-1$, consistent with the experiment (see Supplementary for details of the theoretical method).
Note that these calculations are valid only in the low temperature limit where magnon interactions can be neglected; however, in this regime, they agree with the $\Delta S/N_\phi\approx 0.25k_B$ measured at $T\approx \SI{25}{K}$.
For the lowest temperatures ($T \lesssim \SI{20}{K}$), the measured $\nu = 0$ entropy falls below the theoretical expectation for a Goldstone mode (though in this regime the expected signal is also comparable or smaller than the measurement uncertainty).
This may indicate that the Goldstone mode is weakly gapped by  short-range disorder, which breaks the $U(1)$ valley symmetry\cite{delagrange_vanishing_2024}. Indeed, low-$T$ STM studies observe slow gradients in the phase of the Kekul\'e order that distort near disorder sites~\cite{liu_visualizing_2022}.

\section{Composite fermion entropy: composite fermi liquids and degenerate $\Lambda$-levels}

To further investigate the effect of electronic correlations, we extend the thermodynamic measurements to the partially filled Landau levels. We begin by focusing on the half filled Landau level, where the system remains gapless at low temperature and the entropy remains comparatively large.  Within a composite fermion picture,
\begin{figure}
\centering
\includegraphics[width = 83mm]{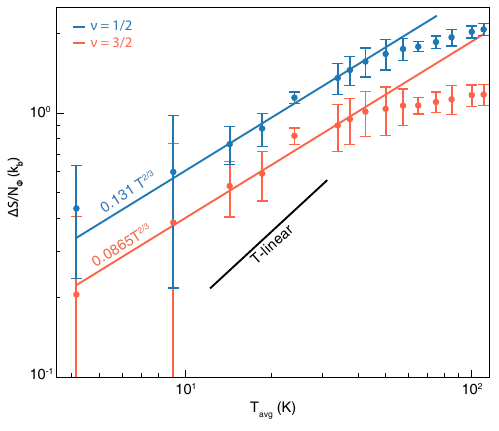}
\caption{Entropy per flux of monolayer graphene half-filled  $N_L=0$ Landau level, which is expected to host a composite Fermi liquid (CFL).
While a Fermi liquid features $T$-linear entropy,  gauge fluctuations in the CFL further contribute,  with $S \propto T^{2/3}$  expected for short-range interactions~\cite{halperin_theory_1993}. While the experimental uncertainty is too large to fit an exponent,  the data is consistent with  faster-than-linear scaling of $S(T)$ at low temperatures. }
\label{fig:power law half filled LL}
\end{figure}
 $\nu \to 1/2$ corresponds to zero magnetic field, with the ground state described by a composite Fermi liquid (CFL)~\cite{halperin_theory_1993}.  As can be seen in Fig.~\ref{fig:1}d, at low-$T$ the $1/2$ states exhibit larger entropy than the integer states, with no sign of a gap.
Indeed, at the mean-field level (when ignoring  gauge fluctuations) the  CF Fermi surface has a density of state per flux quantum $\rho = m_\ast \ell_B^2/\hbar^2 = 1/(x E_C)$, where $x$ is a dimensionless constant that depends on the form of the interaction.
The Fermi surface contributes a linear-$T$ entropy, $S= \frac{\pi^2}{3} k_B T \rho = \frac{\pi^2}{3} k_B T / (x E_C)$. Fig.~\ref{fig:1}e shows the entropy calculated for a single-component CFL treated at this mean-field level
using the CF effective mass estimated in Ref.~\cite{halperin_theory_1993}, $m_\ast \approx 3.3 \hbar^2 / (E_C \ell_B^2) \approx 0.1 m_e$ where $m_e$ is the electron mass.
The magnitude of the evaluated entropy is qualitatively similar to experiment.

Going beyond free CF approximations, gauge fluctuations are expected to both renormalize the CF mass (and thereby the density of states) and contribute additional entropy. As first analyzed by Halperin, Lee, and Read (HLR)~\cite{halperin_theory_1993}, these fluctuations yield an entropy that grows super-linearly as $T\to0$. For a long-range Coulomb interaction, the CF effective mass acquires a logarithmic renormalization—characteristic of a marginal Fermi liquid—resulting in the mean-field prediction $S \propto \frac{k_B T}{E_C}\ln(T_0/T)$. In contrast, for a short-range interaction the fluctuations are stronger, leading to non-Fermi liquid behavior with $S \propto \frac{k_B T}{E_C}(T_0/T)^{1/3} \propto T^{2/3}$ at the mean-field level. In our system, screening from the graphite gates, described by $V(q)=\frac{2\pi}{q}\tanh(qd)$, introduces a new length scale $d$ and prompts a crossover from long-range to short-range scaling as the temperature is lowered. Within the HLR framework, one may infer a crossover temperature of order $k_B T_\ast \sim E_C (\ell_B/d)^2 \sim \SI{3}{K}$; since our lowest measured temperature is $\SI{4}{K}$, we thus expect the entropy to scale between $T^{2/3}$ and $T$. While the calculations presented here are based on mean field theory, we note that calculations based on perturbative expansions also consistently indicate sub-linear scaling of entropy with $T$~\cite{mross_controlled_2010,lee_low-energy_2009,metlitski_quantum_2010}.

In Fig.~\ref{fig:power law half filled LL}, we present a log-log plot of the measured entropy in monolayer graphene at $\nu=1/2$ and $\nu=3/2$, with $T$-linear and $T^{2/3}$ scalings shown for comparison. Although the experimental uncertainty is large and the accessible temperature range is limited, the data suggest that the entropy exceeds a purely $T$-linear behavior and is consistent with $T^{2/3}$ scaling. Future experiments at lower temperatures and with higher precision may enable a definitive quantitative comparison with the $T^{2/3}$ scaling and reveal additional corrections~\cite{lee_low-energy_2009, metlitski_quantum_2010}.

At odd fractional fillings $\nu=p/(2p\pm1)$, the system forms incompressible fractional quantum Hall phases corresponding to integer quantum Hall effects of the composite fermions.  To study these phases in detail, we move away from studying $S/N_\Phi$, which requires referencing our data to a state of fixed entropy.  This fails at sufficiently low temperatures, where the integer quantum Hall gaps become so large that the sample does not charge on timescales of the experiment (practically, the RC time of the sample becomes larger than several minutes).  Instead, we focus on measuring $\mathrm{d}S/\mathrm{d}N$, which can be determined without the need to integration.
 Fig.~\ref{fig:jain_states}a shows $\mu$ in a regime of $\nu$ and applied displacement field $D$ where the valence electrons occupy an $N=0$ orbital state~\cite{hunt_direct_2017,zibrov_tunable_2017} in a bilayer graphene device.
Sharp jumps in $\mu$ correspond to the Jain sequence of incompressible  states~\cite{jain_composite_2005}.
Analyzing the change in $\mu$ between  $T_{\mathrm{hot}} = \SI{600}{mK}$ and $T_{\mathrm{cold}} = \SI{100}{mK}$, we extract $\mathrm{d}S/\mathrm{d}N$, shown in Fig.~\ref{fig:jain_states}b.
$\mathrm{d}S/\mathrm{d}N$ changes sign across the incompressible FQH states, with enhanced entropy for small doping that mimics the integer case of Fig.~\ref{fig:1}.

To understand the entropy in this regime, we consider the thermodynamics of composite fermions in a reduced, density-dependent effective magnetic field $B_*=B(1-2\nu)$.
This problem has previously been studied within the random-phase approximation ~\cite{stern_singularities_1995}, assuming  an unscreened Coulomb interaction and in the absence of disorder. Within this approximation, the jump in the chemical potential at $T \to 0$ (typically referred to as the thermodynamic gap) is predicted to scale as $\Delta \propto \frac{1}{\ln(2p + 1)}$ at $\nu = \frac{p}{2 p + 1}$, due to the renormalization of the CF effective mass. The experimentally observed jumps in $\mu$ fall off much faster than this at $T = \SI{100}{mK}$, as shown in the inset of Fig.~\ref{fig:jain_states}a.
Three effects likely contribute to this discrepancy, including finite temperature, screening of the Coulomb interactions by the gates, and disorder. It is an interesting challenge for theory to account for all three, but we note that the distance to the gates $d_g = \SI{64}{nm}$ is still sizable when compared with the effective CF magnetic length $\ell_B^{\textrm{eff}} = \sqrt{2 p + 1} \ell_B
\approx \SI{21}{nm}$ at $2 p + 1 = 9$, $B = \SI{13.8}{T}$. Similarly, $T = \SI{100}{mK}$ is two orders of magnitude smaller than the gap size where the deviations from the clean theory occur, and little temperature dependence is seen below $\SI{100}{mK}$.

We thus focus first on the effects of disorder. We consider non-interacting CF filling the $\Lambda$ levels, with a phenomenological Gaussian broadening to account for disorder. In this approach, the problem reduces to the thermodynamics of free fermions, and the CF entropy can be obtained via
\begin{equation}
S_{\mathrm{CF}}\Bigl(N_{\mathrm{CF}}=N,B_\ast=B(1-2\nu)\Bigr) = S_e(N,B).
\end{equation}
Figure~\ref{fig:jain_states}b shows the resulting entropy, assuming a CF effective mass of $m_\ast=0.25m_e$ and a Gaussian broadening $\Gamma=\SI{155}{\mu eV}$ (see Supplementary Information for additional details). In the absence of broadening, at $\nu=p/(2p+1)$ the CFs occupy an integer number $p$ of completely filled $\Lambda$ levels and the entropy vanishes in the low-temperature limit. As $\nu$ is detuned from these fillings, the system becomes compressible, and the entropy arises from the motional degrees of freedom of the CFs in a partially occupied $\Lambda$ level, leading to a pronounced sign change in $\mathrm{d}S/\mathrm{d}N$ across the gap. The magnitude of this gap is non-monotonic as a function of $\nu$ due to the competition between two effects. For $\nu=p/(2p+1)$, the addition of one electron introduces $2p+1$ quasiparticles, resulting in a larger entropy per electron as $\nu\to1/2$. Conversely, the CF cyclotron gap scales as $B_\ast\propto (1-2\nu)B$, which eventually suppresses quantum oscillations when disorder becomes comparable to the gap.

Notably, within this simplified model, the entropy is completely dominated by a single $\Lambda$ level for $p=1,2,3$ (see Supplementary Information). In this regime, inter-particle interactions within the nondispersing $\Lambda$ level are expected to play a dominant role in determining the low-temperature entropy. Indeed, discrepancies between experiment and the free-CF model are apparent—for example, between $\nu=1/3$ and $\nu=2/5$—where the experimentally measured entropy is detectably lower than that predicted by the non-interacting model. Due to the Coulomb interaction, the ground state in this regime is expected to form a Wigner crystal of either electrons or fractionally charged quasiparticles, with the residual entropy dominated by density fluctuations (magneto-phonons) of the charge-ordered state. We analyze the entropy in this regime in detail in the next section.

\begin{figure}[ht]
    \centering
    \includegraphics[width = \columnwidth]{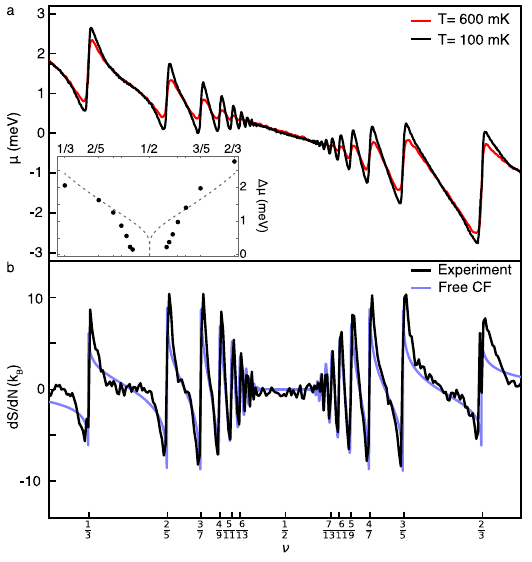}
    \caption{\textbf{Entropy in the fractional quantum Hall regime of bilayer graphene.}
     \textbf{(a)} Chemical potential $\mu$ in the partially-filled $N_L = 0$ Landau level at $B = \SI{13.8}{T}$ at $T = \SI{600}{\milli\kelvin}$ (red) and $T = \SI{100}{\milli\kelvin}$ (black).  Inset: Thermodynamic energy gaps determined from the chemical potential jumps at 100mK.  Dashed line is a guide for the eye showing expected $1/\ln (2p+1)$ scaling.
    \textbf{(b)} Entropy per electron $\mathrm{d}S/\mathrm{d}N$ obtained from the data in panel (a) (black curve).
    Also shown (in blue) is the calculated entropy from a non-interacting composite fermion model at $T=\SI{350}{mK}$, assuming a Gaussian Landau level broadening $\Gamma=150 \mu eV$.}
    \label{fig:jain_states}
\end{figure}

\section{Integer and fractional Wigner crystals}

\begin{figure}[ht]
    \centering
    \includegraphics[width = \columnwidth]{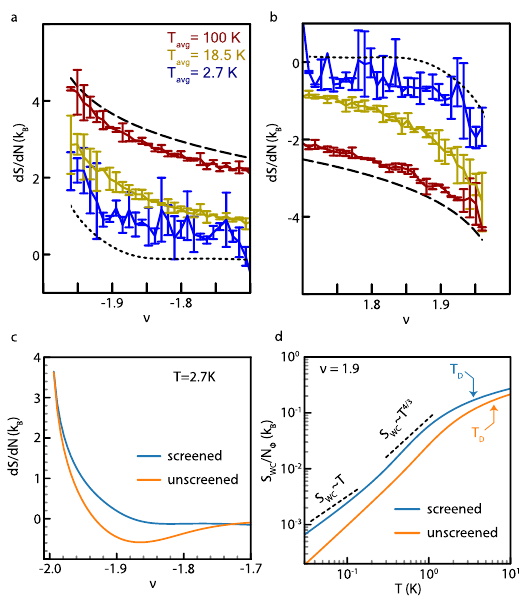}
    \caption{\textbf{Entropy per electron near integer fillings in monolayer graphene.}
    {\bf (a)} $\nu$-dependence of the entropy per added particle, $\mathrm{d}S/\mathrm{d}N$, near $\nu = -2$ and {\bf (b)} $\nu=2$  at various temperatures. The dashed curves indicate the high-$T$ motional entropy of a four-flavor degenerate Landau level, while the dotted curve shows the entropy predicted by a classical Wigner crystal (WC) model at $T = \SI{2.7}{K}$. The three temperature ranges are: $T_{\mathrm{hot}} = \SI{120}{K}$ to $T_{\mathrm{cold}} = \SI{80}{K}$; $T_{\mathrm{hot}} = \SI{20.9}{K}$ to $T_{\mathrm{cold}} = \SI{16.1}{K}$; and $T_{\mathrm{hot}} = \SI{3.4}{K}$ to $T_{\mathrm{cold}} = \SI{2}{K}$.
    {\bf (c)} Calculated entropy as a function of temperature for an electron Wigner crystal at $\nu = -1.9$, computed with and without screening. For $T \lesssim T_D$, $\mathrm{d}S/\mathrm{d}N < 0$ due to WC stiffening with increasing density. This regime is more challenging to access when screening is present. {\bf (d)} Theoretical predictions for $\mathrm{d}S/\mathrm{d}N$ at $T = \SI{2.7}{K}$ are shown for an unscreened Coulomb interaction and for one screened by metallic gates and Landau-level mixing.
    }
    \label{fig:experiment_WC_entropy_integer}
\end{figure}
In the vicinity of both integer and fractional quantum Hall states, quasiparticles emerge with density proportional to the deviation of $\nu$ from the nearest quantized value, i.e.\ $n_{\text{qp}} \propto \nu - \nu_0$ where $\nu_0$ corresponds to an incompressible integer or fractional Hall state. We first examine the behavior of $\mathrm{d}S/\mathrm{d}N$ near $\nu_0 = \pm 2$ (see Fig.~\ref{fig:experiment_WC_entropy_integer}). At high temperatures, the motional entropy of these quasiparticles predominates. As shown in Figs.~\ref{fig:experiment_WC_entropy_integer}a--b, the dashed curves represent the high-$T$ theoretical result computed via Eq.~\eqref{eq:1dome} from the motional entropy of a single Landau level, and the experimental data closely follow this prediction. At high $T$, the reduction in $\mathrm{d}S/\mathrm{d}N$ with increasing $n_{\text{qp}}$ arises from the indistinguishability of electrons and holes, since the statistical factor $1/N!$ reduces the effective number of accessible states.

As the temperature is lowered, $\mathrm{d}S/\mathrm{d}N$ decreases as interactions and disorder progressively freeze quasiparticle motion. The shape of $\mathrm{d}S/\mathrm{d}N$ versus $\nu$ also changes, indicating a distinct state at low $T$.  In the absence of disorder and at low densities and temperatures, we expect the quasiparticles to form a triangular-lattice Wigner crystal (WC)~\cite{fukuyama_two-dimensional_1975, platzman_quantum_1993, pan_transition_2002, kumar_unconventional_2018, kim_quantum_2021}, as has been recently visualized via scanning probe microscopy measurements in bilayer graphene~\cite{tsui_direct_2024}.
Within the WC regime, the decrease in $\mathrm{d}S/\mathrm{d}N$ with increasing $n_{\text{qp}}$ is attributed to the behavior of the magneto-phonon modes. At low $n_{\text{qp}}$, weak interactions yield soft magneto-phonon modes and a high entropy per particle $S/N$. As $n_{\text{qp}}$ increases, strengthening interactions stiffen the WC, shifting the magneto-phonon modes to higher energies and reducing $S/N$. The sign and magnitude of $\mathrm{d}S/\mathrm{d}N$ are determined by the competition between the increased density of modes and the reduced entropy per mode; when the latter dominates (as at higher $n_{\text{qp}}$), $\mathrm{d}S/\mathrm{d}N$ decreases and eventually reverses sign. This behavior is captured by our theoretical model, as indicated by the dotted curves in Figs.~\ref{fig:experiment_WC_entropy_integer}a--b, which depict the predicted entropy of an electron WC at $T = \SI{2.7}{K}$.

We now briefly describe the theoretical model for the entropy of the WC. Building on previous studies of magneto-phonon modes~\cite{fukuyama_two-dimensional_1975, giamarchi_disordered_2002}, we compute the entropy $S$ from the phonon dispersion $\epsilon(\mathbf{k})$ by treating the phonon excitations as free bosons, akin to the Debye model. Although we expect the WC to melt in the dilute limit---particularly in the fractional regime where energy scales are low---Ref.~\onlinecite{yang_thermopower_2009} shows that magneto-phonons remain the dominant source of entropy even above the melting temperature. Moreover, the same framework applies to a WC of fractional quasiparticles by replacing the elementary charge $e$ with the fractional charge $e^*$ and the magnetic length $\ell_B$ with $\ell_B^* = \sqrt{\hbar/(e^*B)}$, assuming exchange statistics are negligible in a localized crystal~\cite{cooper_observable_2009, yang_thermopower_2009}.

In the case of an unscreened Coulomb interaction, the high-field WC entropy is predicted to scale as \cite{cooper_observable_2009}
\begin{equation}
    \frac{S_{\mathrm{WC}}}{N_\phi} \propto \left(\frac{T}{T_D(\nu)}\right)^{4/3},
\end{equation}
where $T_D(\nu) \sim E_C \sqrt{e/e^*} |\nu - \nu_0|^{3/2}$. We find that this formula underestimates the WC entropy by an order of magnitude at relevant temperatures. To calculate the WC entropy more quantitatively, we adopt a WC model that incorporates three screening effects; notably, our model allows us to quantitatively account for the behavior of $\mu$ at low temperatures
\cite{assouline_energy_2024}.
First, screening from the dual graphite gates at a distance $d_g$ renders the $q=0$ portion of the Coulomb interaction finite. Consequently, the magneto-phonon dispersion crosses over from $\epsilon(k) \propto k^{3/2}$~\cite{fukuyama_two-dimensional_1975, giamarchi_disordered_2002} for $k d_g \gtrsim 1$ to $\epsilon(k) \propto k^2$ for $k d_g \lesssim 1$. This crossover causes the magneto-phonon entropy $S_{\text{WC}}(T,\nu)$ to change from a $T^{4/3}$ to a linear-$T$ dependence at the lowest temperatures, as shown in Fig.~\ref{fig:experiment_WC_entropy_integer}d. Second, strong Landau-level (LL) mixing in monolayer and bilayer graphene, attributable to the filled Dirac sea, softens the interactions and lowers the Debye temperature $T_D$. We include LL mixing at the level of the static random-phase approximation~\cite{shizuya_electromagnetic_2007, misumi_electromagnetic_2008, gorbar_dynamics_2010, papic_topological_2014, zibrov_tunable_2017}. Third, when applied to fractional quantum Hall liquids with small gaps, we model the static dielectric response of the FQH liquid as $\Pi_{\text{FQH}}(q) = -b q^2 + \cdots$, where $b$ is a fitting parameter determined by comparing to experimental data (see Fig.~\ref{fig:WC_theory_v_exp}a)~\cite{assouline_energy_2024}, though its precise value plays a minor role in the resulting entropy.

Altogether, the screened interaction is given by
\begin{equation}
    V_{\text{scr}}(q) = \frac{V(q)}{1 - \Pi(q)V(q)},
\end{equation}
where $V(q) = E_c \ell_B \frac{2\pi}{q}\tanh(q d_g)$, and $\Pi(q)$ is either the non-interacting polarization $\Pi_{\text{LL}}(q)$ due to inter-LL transitions for graphene~\cite{shizuya_electromagnetic_2007, misumi_electromagnetic_2008, gorbar_dynamics_2010, papic_topological_2014, zibrov_tunable_2017} or $\Pi_{\text{FQH}}(q)$ for FQH Wigner crystals (see Supplementary for the precise form of $\Pi_{\text{LL}}(q)$ and $\Pi_{\text{FQH}}(q)$). In Figs.~\ref{fig:experiment_WC_entropy_integer}c--d, we compare the calculated $\mathrm{d}S/\mathrm{d}N$ near $\nu = -2$ with and without accounting for these screening effects. As shown in Fig.~\ref{fig:experiment_WC_entropy_integer}c, incorporating screening is essential to capture the experimentally observed trends. In particular, screening reduces $T_D$ and therefore raises the entropy.

Finally, we comment on the physical meaning of $\mathrm{d}S/\mathrm{d}N$ and how it can be used to infer $T/T_D$. At or above the Debye temperature, $\mathrm{d}S/\mathrm{d}N > 0$ because the system follows equipartition and so the entropy follows the density of magneto-phonon modes, which increases with $N$. At $T \ll T_D$, however, these modes partially freeze out.  In this case, $\mathrm{d}S/\mathrm{d}N < 0$ for large $n_{\text{qp}}$ because the Wigner crystal becomes more rigid with increasing $N$, supressing the entropy. Including screening brings $T = \SI{2.7}{K}$ close to $T_D$, making both theory and experiment yield $\mathrm{d}S/\mathrm{d}N \approx 0$. As illustrated in Fig.~\ref{fig:experiment_WC_entropy_integer}d, screening also makes it more challenging to reach the $T < T_D$ regime. Currently, the limited conductivity of the WC phase restricts the achievable charging rates at the lowest temperatures.
Future IQHE experiments that use alternative device geometries or charging methods may reach the $T < \SI{1}{K}$ regime. If observed, $\mathrm{d}S/\mathrm{d}N < 0$ would provide further confirmation for the Wigner crystal interpretation.

\subsection{Anyonic Wigner crystals and the even-denominator state}

Obtaining a quantitative understanding of the entropy of dilute fractional quantum Hall (FQH) quasiparticles is especially important for experimentally identifying non-Abelian anyons via thermodynamic measurements. As discussed in the introduction, a collection of $N_a$ well-separated non-Abelian anyons is predicted to contribute a topological entropy
\begin{align}
S_{\textrm{topo}} = N_a k_B \ln(\mathcal{D}),
\end{align}
where $\mathcal{D} = \sqrt{2}$ for the Ising anyons anticipated near filling $\nu_0 = -1 + 1/2$ in bilayer graphene~\cite{papic_topological_2014, ki_observation_2014,zibrov_tunable_2017,li_even_2017}. Taking $\mathrm{d}N_a/\mathrm{d}N = \pm4$ for quasiparticles of charge $\pm e/4$ then implies
\begin{equation}
\frac{\mathrm{d}S_{\textrm{topo}}}{\mathrm{d}N} = 4 k_B \ln \sqrt{2} \; \text{sgn}(\nu - \nu_0),
\end{equation}
which is independent of temperature~\cite{cooper_observable_2009}.

This topological contribution arises \emph{in addition} to the motional entropy of the anyons. Ideally, the motional entropy freezes out at low temperatures, leaving only $S_{\textrm{topo}}$. In practice, however, whether there exists an accessible regime where $S_{\textrm{topo}}$ dominates the measured entropy is subtle. For example, at sufficiently low quasiparticle density $N_a$, the motional entropy may remain significant at experimental temperatures, concealing the topological contribution. Conversely, at higher quasiparticle densities, inter-anyon interactions can lift the topological degeneracy and entirely suppress the topological entropy. In the extreme limit, these interactions can drive the formation of new incompressible fractional quantum Hall ``daughter'' states at sufficiently high $N_a$~\cite{levin_collective_2009}.

\begin{figure}[ht]
    \centering
    \includegraphics[width = \columnwidth]{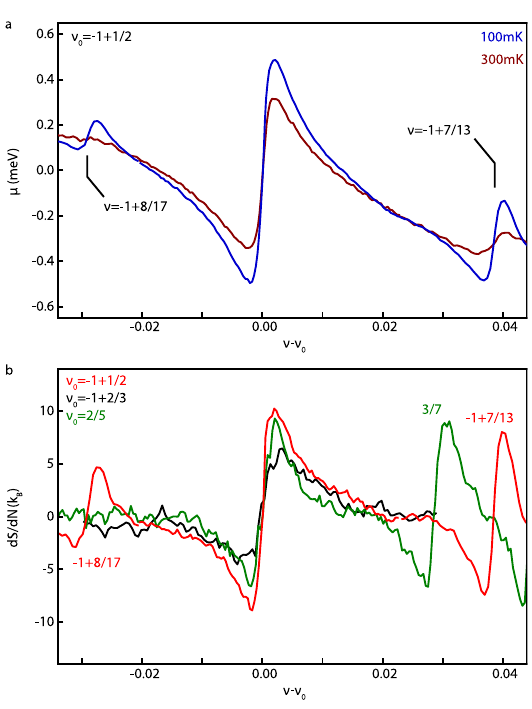}
    \caption{
    \textbf{Entropy of the bilayer graphene $\nu=-1 + 1/2, -1 + 2/3$ and $2/5$ states at $B=\SI{13.8}{T}$.}
    \textbf{(a)} Measured $\mu$ at $T=\SI{300}{mK}$ and $T=\SI{100}{mK}$ near $\nu_0=-1/2$ at $B=13.8T$.
    \textbf{(b)} Measured $\mathrm{d}S/\mathrm{d}N$ near $\nu_0=-1/2$, $-1/3$, and $2/5$ extracted from the measured $\mu$ using $T_{\mathrm{hot}} = \SI{300}{mK}$ to $T_{\mathrm{cold}} = \SI{100}{mK}$. }
         \label{fig:FQH_WC_exp}
\end{figure}

Fig.~\ref{fig:FQH_WC_exp}a shows the chemical potential $\mu$ measured near $\nu_0 = -1+1/2$ in bilayer graphene at $T = \SI{100}{mK}$ and $\SI{300}{mK}$. Figure~\ref{fig:FQH_WC_exp}b displays the corresponding $\mathrm{d}S/\mathrm{d}N$ extracted from these measurements, alongside data for $\nu_0 = -1 + 2/3$ and $\nu_0 = 2/5$ in the same temperature range. These two additional states, with anticipated quasiparticle charges $\pm e/3$ and $\pm e/5$, respectively, offer useful comparisons to the $\pm e/4$ excitations of the $\nu=-1+1/2$ state. The measured thermodynamic gaps are also comparable: $\Delta \mu_{-1+2/3} = \SI{1.3}{meV}$, $\Delta \mu_{-1+1/2} = \SI{1.0}{meV}$, and $\Delta \mu_{2/5} = \SI{1.5}{meV}$.

\begin{figure*}[ht]
    \centering
    \includegraphics[width = \textwidth]{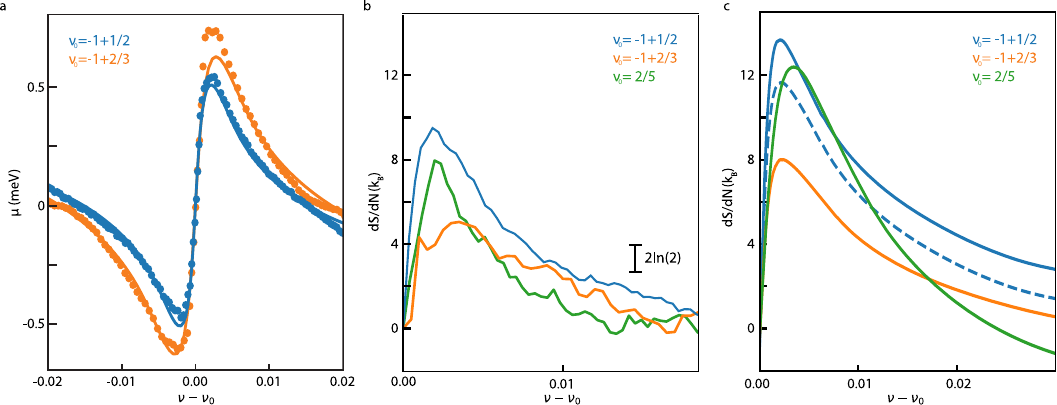}
    \caption{
    \textbf{Comparison of theoretical bilayer graphene FQH Wigner model with experiment.}
    \textbf{(a)} Wigner crystal model fit to the $T = \SI{100}{mK}$ chemical potential data, from which model parameters—including the gap size $\Delta$, screening parameter $b$, and disorder broadening $\Gamma$—were extracted. Dots indicate the experimental data, and solid lines represent the model fit.
    \textbf{(b)} Experimental $\mathrm{d}S/\mathrm{d}N$ data from Fig.~\ref{fig:FQH_WC_exp}, anti-symmetrized about $\nu_0$.
    \textbf{c} Theoretically predicted Wigner crystal entropy in the vicinity of different fractional fillings. For the $\nu_0=-1+1/2$ curve, the non-Abelian contribution to the entropy~\cite{cooper_observable_2009} is either accounted for (solid), or neglected (dashed).}
         \label{fig:WC_theory_v_exp}
\end{figure*}

As with conventional FQH phases in Fig.~\ref{fig:jain_states}, a sizable $\mathrm{d}S/\mathrm{d}N$ appears near $\nu_0$, reflecting residual motional entropy of the low-density quasiparticle gas. Close to the $\nu = -1 + 1/2$ state, large $\mathrm{d}S/\mathrm{d}N$ also appears in the vicinity of
$\nu=-1+8/17$ and $\nu=-1+7/13$, consistent with the formation of Levin--Halperin fractional quantum Hall
states~\cite{levin_collective_2009}. These states emerge as interacting liquids of the $e/4$ quasiparticles from the $\nu=-1+1/2$ state and thus mark an upper limit on the anyon density at which the topological entropy can be detected. Similarly, the presence of the 3/7 state sets a limit on how many charge-$e/5$ anyons can be added to the $2/5$ state before their interactions drive the formation of a new incompressible phase with entirely different elementary excitations. Consequently, if a density range exists where the topological entropy $S_{\textrm{topo}}$ dominates, it would likely occur in the window $0.005 < \lvert \nu - \nu_0 \rvert < 0.02$, where Coulomb repulsion can localize anyons into a Wigner crystal without fully lifting the topological degeneracy.

Over the temperature range $T = 100\text{--}300\,\mathrm{mK}$, the $\nu = -1/2$ data do not show a clear plateau at $S_{\mathrm{topo}}$. Nonetheless, $\mathrm{d}S/\mathrm{d}N$ near $\nu = -1/2$ exceeds that of both $\nu = -1 + 2/3$ and $\nu = 2/5$, remaining above $S_{\mathrm{topo}}$ for all $\lvert \nu - \nu_0 \rvert < 0.016$. These observations leave open the possibility of detecting non-Abelian entropy at lower temperatures, while indicating that motional entropy still plays a significant role under current experimental conditions. In line with the integer WC results, we observe $\mathrm{d}S/\mathrm{d}N \gtrsim 0$ in the Abelian states and $\mathrm{d}S/\mathrm{d}N > dS_{\mathrm{topo}}/dN$ near $\nu = - 1/2$, suggesting that $T \gtrsim T_D$. Reaching $T < T_D$ may thus be necessary to fully isolate $S_{\mathrm{topo}}$, which would provide an unambiguous thermodynamic signature of non-Abelian anyons.

\begin{figure*}[ht]
\centering
\includegraphics[width = \textwidth]{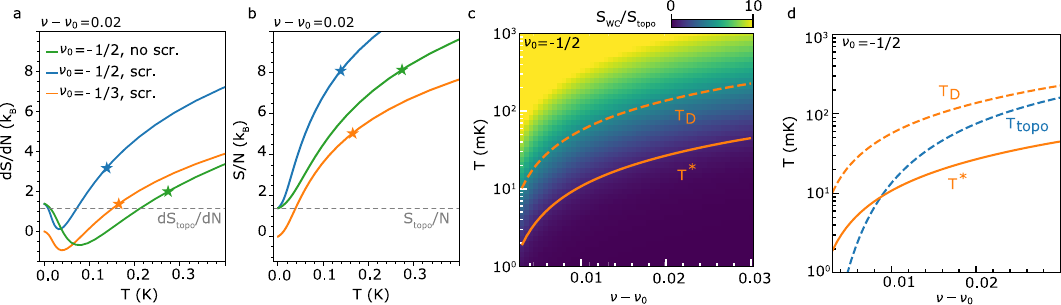}
\caption{\textbf{Theoretically determined Wigner crystal entropy for bilayer graphene at B = 13.8T}.
\textbf{(a)} Temperature dependent $\mathrm{d}S/\mathrm{d}N$ and \textbf{(b)}  $S/N$ for $\nu-\nu_0=0.02$ near $\nu_0=-1/2$ and $-1/3$ in different theoretical models described in the main text.
The $\star$ symbols indicate the Debye temperature $T_D$ for each curve, while the horizontal dashed line represents the expected topological contribution near $\nu_0 =- 1/2$.
\textbf{(c)} Ratio of Wigner crystal entropy $S_{\mathrm{WC}}$ and  $S_{\mathrm{topo}}$,
plotted over the experimentally relevant range of $\nu-\nu_0$ and $T$.
$T^*$, defined as the temperature where $S_{WC}=S_{topo}$, as well as $T_D$, are shown in overlay.
\textbf{(d)} Comparison of $T_D$, $T^*$, and $T_{topo}$, the scale characterizing splitting of the  topological degeneracy by anyon interactions. The desired $T_{\mathrm{topo}}<T<T^*$ regime is indicated by the green shading.
}
\label{fig:theory_FQH_WC_TDep}
\end{figure*}

In order to assess whether future experiments at lower temperatures and higher sensitivity could isolate $S_{\mathrm{topo}}$, we compare our results with theoretical modeling of $S(\nu, T)$. In the vicinity of the $\nu_0 = -1 + 1/2, -1 + 2/3,$ and $2/5$ states, we assume fractional quasiparticles crystallize into a triangular lattice WC, enabling us to compute the entropy via the framework outlined in the previous section. For $\nu - \nu_0 = 0.01$ and $B = \SI{14}{T}$, the lattice spacing $a$ reaches $\sim 13\,\ell_B \approx 90\,\mathrm{nm}$, where exchange effects are negligible~\cite{cooper_observable_2009, yang_thermopower_2009}. At higher densities $\nu - \nu_0 \gtrsim 0.02$, exchange effects become non-negligible, leading to the formation of new FQH liquids; the WC calculation is thus not expected to hold in this regime. At lower densities, disorder must be incorporated into the model to quantitatively capture the behavior.
We identify three distinct ways in which disorder influences $\mu(\nu)$ and, consequently, $\mathrm{d}S/\mathrm{d}N$. First, the long-wavelength component of the disorder potential causes inhomogeneous broadening: even within the gap, $\pm e^*$ excitations exist in the system, thus rounding the jump in $\mu$. We account for this broadening in the same manner as Ref.~\onlinecite{assouline_energy_2024}, treating its width $\Gamma$ as a phenomenological parameter. By fitting $\mu(T=\SI{100}{mK},\nu)$ near $\nu_0=-1+1/2, -1+2/3$ to the fractional Wigner crystal model, we fix $\Delta, b,$ and $\Gamma$ as shown in Fig.~\ref{fig:WC_theory_v_exp}a. The fitting procedure and resulting parameters are discussed in the Supplementary materials. Since the same bilayer graphene device was used in Ref.~\onlinecite{assouline_energy_2024}, we adopt identical parameters here. Next, the inhomogeneous broadening also complicates the extraction of $S_{\mathrm{topo}}$ and $\mathcal{D}$. The total density change $\mathrm{d}N$ reflects the difference between the number of $+e^*$ and $-e^*$ anyons, whereas $S_{\mathrm{topo}}$ depends on their sum. Consequently, if disorder induces quasiparticles in some regions and quasiholes in others, $\Delta N$ may not track the true change in the total number of excitations $N_a$. This mismatch remains negligible for $\lvert \nu - \nu_0\rvert \gtrsim 0.005$ but can matter closer to $\nu_0$ (See Supplementary materials). Finally, short-range disorder can pin the Wigner crystal, generating a ``pinning gap'' $T_{\mathrm{pin}}$~\cite{giamarchi_disordered_2002}. In principle, for $T<T_{\mathrm{pin}}$, the WC would be immobilized and its entropy strongly reduced. However, lacking a microscopic understanding of the disorder distribution and any direct experimental evidence for such a gap, we adopt the minimal assumption $T_{\mathrm{pin}}=0$.

With the theoretical parameters thus fixed by experiment, we may compare the predicted $\mathrm{d}S/\mathrm{d}N$, shown in Fig.~\ref{fig:WC_theory_v_exp}c, with the experimental data in Fig.~\ref{fig:WC_theory_v_exp}b. To facilitate this comparison, we anti-symmetrize the experimental $\mathrm{d}S/\mathrm{d}N$ with respect to a particle--hole transformation about each $\nu_0$. We find overall qualitative agreement between theory and experiment: the relative magnitudes of $\mathrm{d}S/\mathrm{d}N$ near the $\nu_0 = -1/3, 2/5,$ and $-1/2$ states match well, the predicted crossing of the $-1/3$ and $2/5$ curves is observed, and $\mathrm{d}S/\mathrm{d}N$ remains positive throughout the range examined. Figure~\ref{fig:WC_theory_v_exp}c also shows theoretical results for $\nu_0 = -1/2$ both with and without the topological contribution $dS_{\mathrm{topo}}/dN$. As the experimental data lack the requisite precision to distinguish these two cases, we cannot definitively conclude whether $S_{\mathrm{topo}}$ has been observed.

Our benchmarked theoretical model can be used to evaluate how future experiments might unambiguously detect $S_{\mathrm{topo}}$. Figs. ~\ref{fig:theory_FQH_WC_TDep}a-b show $dS/dN$ and $S/N$ computed for $\nu-\nu_0=0.02$ for $\nu_0=-1/2$ with and without screening effects and in the absence of disorder.  Similar calculations for $\nu_0=1/3$ are also shown for comparison.  Screening reduces $T_D$ by roughly a factor of two.  As a result, although $T_D$ provides a useful baseline to estimate the temperatures required to suppress WC excitations, more refined criteria can be developed. A useful temperature scale is defined by  $S_{\mathrm{WC}}(T^*) =S_{\mathrm{topo}}$; below this temperature, the topological entropy comprises the majority of the total entropy. Measurements in the $T<T^*$ regime might hope to infer $S_{topo}$ by extrapolationn.
As shown in Fig. \ref{fig:theory_FQH_WC_TDep}c, empirically $T_*\approx 0.1T_D$, well below 100mK for the experimentally relevant range.

Even if temperatures below $T^*$ become experimentally accessible at moderate $\nu-\nu_0$, one encounters a second constraint: at larger $\nu-\nu_0$, the quasiparticle interaction energy may split the topological degeneracy, introducing an additional scale $T_{\mathrm{topo}}$. To observe $S_{\mathrm{topo}}$, the experiment must then satisfy $T_{\mathrm{topo}} < T < T^*$. Although $T_{\mathrm{topo}}$ has not been directly measured, theoretical estimates based on model wave functions and unscreened Coulomb interactions suggest $k_B T_{\mathrm{topo}} \sim 0.01 E_C  e^{-a/\xi},$ where $E_C = e^2/(4\pi\epsilon\ell_B)$ is the Coulomb scale, $a$ is the inter-anyon spacing, and $\xi\approx 2.3\,\ell_B$ is the neutral-fermion correlation length~\cite{baraban_numerical_2009}. As shown in Fig.~\ref{fig:theory_FQH_WC_TDep}d, this estimate indicates $T_{\mathrm{topo}} < T^*$ only for $\lvert\nu-\nu_0\rvert < 0.01,$ where $T^*$ can fall below $\SI{10}{mK}$.
While these numbers appear challenging for near-term experiments, the same screening effects that reduces the WC energy scale very likely lower $T_{\mathrm{topo}}$ as well.
However, the quantitative impact of screening on $T_{\mathrm{topo}}$ is not well established, and systematic numerical studies of realistic interaction potentials are needed to clarify the likely size of the temperature window.

Looking ahead, the primary obstacle to accessing the low-T regime in the current experiment is the resolution with which $\mu$ can be determined. Since entropy is obtained by differentiating the chemical potential with respect to temperature, the temperature difference $\delta T$ must be large enough to produce a detectable change $\delta \mu$ that exceeds both systematic and random noise. This requirement sets a minimum $\delta T$ of about 200\,mK in our measurements. Overcoming this limitation will likely demand a more sensitive transducer than the graphene sensor layer; for instance, single-electron transistors, with input voltage noise in the $n\text{V}/\sqrt{\text{Hz}}$ range~\cite{zimmerli_voltage_1992}, may enable sufficiently precise detection of $\mu$.
One can also consider modifications to the sample which might help quench the motional entropy. Disorder might accomplish this by generating a pinning gap $T_{\mathrm{pin}}$, which our theoretical modeling has neglected. Alternatively, fabricating a small quantum dot could allow a single non-Abelian anyon to be localized~\cite{ben-shach_detecting_2013}. Our present modeling indicates that, at least in an idealized (disorder-free) system, the energy scales needed to suppress Wigner-crystal entropy remain experimentally challenging, thereby motivating further studies of deliberate confinement.

\section{Conclusion}
We have explored the entropy of monolayer and bilayer graphene in the quantum Hall regime across four orders of magnitude in temperature, ranging from \SI{200}{mK} to \SI{300}{K}. The entropy exhibits a clear energy hierarchy, with different physical phenomena dominating at distinct temperature scales. At \SI{100}{K}, the occupational entropy of four-flavor electrons is dominant, forming a characteristic dome across the zero energy Landau level. As the temperature decreases, various correlated states emerge in sequence. At $T\approx \SI{50}{\kelvin}$, quantum Hall ferromagnetism becomes dominant, splitting the zero energy Landau level into four distinct domes arising from the motional entropy of single-flavor electrons. At integer fillings in this temperature regime, thermally excited magnons become the primary contributors to the entropy, whose detailed temperature dependence provides  entropic evidence for Kekul\'e or anti-ferromagnetic order at $\nu = 0$. As the temperature is further reduced to $T\approx \SI{1}{K}$, charge order begins to manifest at small doping levels relative to integer fillings, indicated by deviations in the filling dependence from that of the full configurational entropy. The fractional quantum Hall states are revealed as dips in the entropy at fractional fillings. Using a free composite fermion model, we qualitatively reproduce the observed entropy patterns. To probe the entropy associated with the quantum degeneracy of non-Abelian anyons, we focus on the entropy per electron at finite doping near the Pfaffian state at $\nu = -1/2$. At low doping, the entropy is dominated by density modes, while at higher doping, we observe excess entropy near $\nu = -1/2$ compared to the Jain fractions. We interpret this excess as a potential signal of the entropy associated with non-Abelian anyons. However, more precise measurements at lower temperatures and with higher sensitivity are needed to confirm this interpretation.

\textbf{Acknowledgments.}
The authors would like to acknowledge discussions with Ady Stern and Bertrand Halperin.
The work was primarily supported by the U.S. Department of Energy,
Office of Science, National Quantum Information Science Research Centers, Quantum Science Center.
T.W. and M.Z. are supported by the U.S. Department of Energy, Office of Science, Office of Basic Energy Sciences, Materials Sciences and Engineering Division under Contract No. DE-AC02-05-CH11231 (Theory of Materials program KC2301).
T.W. is also supported by the Heising-Simons Foundation, the Simons Foundation, and NSF grant No. PHY-2309135 to the Kavli Institute for Theoretical Physics (KITP).
R.F. is
supported by the Gordon and Betty Moore Foundation.
K.W. and T.T. acknowledge support from the Elemental Strategy Initiative conducted by the MEXT, Japan (Grant Number JPMXP0112101001) and JSPSKAKENHI (Grant Numbers 19H05790, 20H00354 and 21H05233). Additional support was provided by a Brown Investigator Award to AFY.  A portion of this work was performed in the UCSB Nanofabrication Facility, an open access laboratory.

 \bibliographystyle{science}
\bibliography{references_andrea}


\clearpage
\newpage
\pagebreak

\onecolumngrid

\begin{center}
\textbf{\large Supplementary information }\\[5pt]
\end{center}

\setcounter{equation}{0}
\setcounter{figure}{0}
\setcounter{table}{0}
\setcounter{page}{1}
\setcounter{section}{0}
\makeatletter
\renewcommand{\theequation}{S\arabic{equation}}
\renewcommand{\thefigure}{S\arabic{figure}}
\renewcommand{\thepage}{\arabic{page}}

\section{Materials and Methods}

\textbf{Sample preparation}

Fabrication details for the monolayer chemical potential sample is given in Ref.~\cite{yang_experimental_2021} and for the bilayer sample in Ref.~\cite{assouline_energy_2024}.
\\

\textbf{Slow charging and LED illumination}

When the monolayer graphene sample is tuned into an incompressible quantum Hall state, it requires more time to charge due to the significantly suppressed conductance between the contacts and the device bulk. This delay impedes precise chemical potential measurements while sweeping the top and bottom gates, as it causes hysteresis in the conductance minima or $\mu$ measured by the detector layer. Fig.~\ref{fig:s1}a shows the change of the top gate voltage over time, while the sample layer is fixed at $\nu=-5/3$ and the top gate is continuously adjusted to set the detector layer at its conductance minimum. If the charging rate is fast, then the top gate does not need to be adjusted over time once the detector layer is tuned at its conductance minimum. However, we found that a slow change in the top gate voltage is needed to maintain the detector layer at the conductance minimum, owing to the long equilibration time of the monolayer graphene sample.

To address the very slow charging issue of the monolayer graphene sample, we illuminate the dual gated graphene device with a LED and accelerate the charging rate through photoinduced conduction at each sample layer filling factor. After illumination, we wait for 10sec $\sim$ 30sec to allow the electrons to cool down before measuring the conductance minimum in the detector layer. Fig.~\ref{fig:s1} shows the effect of LED illumination on the charging rate. Without LED illumination, we observe hysteresis in $\mu$ when sweeping $\nu$ in opposite directions. This hysteresis suggests that the electrons in the sample layer has not reached equilibrium when sweeping $\nu$. On the other hand, under LED illumination described above, hysteresis in $\mu$ disappears, indicating that the charging issue is resolved with illumination.
\\

\textbf{Entropy measurements}

In order to determine the entropy using Maxwell relation, $\rm |\frac{d\mu}{dT}|_N=-|\frac{dS}{dN}|_T$, we measured the chemical potentials at different temperatures. In each chemical potential curve in Fig.~\ref{fig:1}a, we calibrated the constant offsets in $\nu$ and $\mu$ to account for changes in chemical potential in the detector layer with temperature. In the main text Fig.~\ref{fig:1}, Fig.~\ref{fig:power law half filled LL} and Fig.~\ref{fig:experiment_WC_entropy_integer}, these offsets are calibrated by assuming particle-hole symmetry in the lowest Landau level of monolayer graphene: $\mu(\nu\geq0)\approx-\mu(\nu\leq0)$. For the fractional quantum Hall states of bilayer graphene in Fig.~\ref{fig:jain_states}, Fig.~\ref{fig:FQH_WC_exp} and Fig.~\ref{fig:WC_theory_v_exp}, the offset in $\nu$ is calibrated by aligning the sharp incompressible peak at the same fractional quantum Hall states at two different temperatures. For Fig.~\ref{fig:jain_states} and Fig.~\ref{fig:FQH_WC_exp}, a density independent offset in $\mathrm{d}S/\mathrm{d}N$ is applied arbitrarily. These arbritary offsets cancel when plotting the anti-symmetrized entropy per particle in Fig.~\ref{fig:WC_theory_v_exp}.
\\

\textbf{Temperature range for applying $S_0 = S(\nu \approx \pm2)$}

In Fig.~\ref{fig:1} of the main text, the relative change of entropy $\Delta S$ is referenced to the average value of $S_{0}$ measured at $\nu = -1.97$ and +1.97 to calibrate the offset that arises from the integration of $dS/dN$ since $S$ is expected to be zero in a fully filled or empty Landau level at low temperature. However, this method cannot be applied when $S$ becomes nonzero around $\nu =$ -2 and +2, particularly at high temperatures, as thermal excitations of electrons/holes lead to an increase in entropy in the incompressible states. To estimate the temperature range over which the method remains valid, we examined the temperature dependence of entropy between 4.2K and 250K. Fig.~\ref{fig:s2} shows that the entropy at $\nu=0$ reaches its maximum value between 120K and 160K and then decreases as $T_{avg}$ increases (see the green line). The decrease in entropy observed at $T_{avg} \gtrsim$ 120K is attributed to an increase in $S_{0}$, from which $S$ is subtracted. To support this interpretation, we compared $\Delta S$ at $\nu= -1.97$ with $S$ calculated from the free fermion model (see the blue line). The free fermion model indicates that $S$ remains zero at $T_{avg} <$ 120K and begins to increase at a higher temperature. On the other hand, the measured values of $\Delta S$ at $\nu=-1.97$ remain near zero upto the highest temperature 250K. This discrepancy between the calculated and measured entropy suggests that $S_0$ cannot be used as a baseline for correcting the integration offset at $T_{avg} \gg$ 120K.
\\

\textbf{Estimation of error bars for $\Delta S$ and $\mathrm{d}S/\mathrm{d}N$}

In main text Fig.~\ref{fig:1}, the main sources of error in $\Delta S= S - S_{0}$ are the integration offset error and the particle-hole asymmetry. First, as described in "Temperature range for applying $S_0 = S(\nu \approx \pm2)$", the average value of $S$ at $\nu \pm 1.97$ is used to calibrate the integration offset. We referenced the entropy at $|\nu|=1.97$ rather than $|\nu|=2$ because large errors in $\mathrm{d}S/\mathrm{d}N$ occur near $\nu \pm 2$ due to misalignment in $\nu$ between the two chemical potential curves used in calculating $\mathrm{d}S/\mathrm{d}N$. However, referencing the entropy at $|\nu|=1.97$ omits a portion of the entropy between $\nu=-2$ and $\nu=-1.97$, as well as $\nu=1.97$ and $\nu=2$. To estimate the uncertainty in the integration offset resulted from this omission, we evaluated  the integrals $\int_{-1.995}^{-1.97} \frac{dS}{dN} \, \mathrm{d}\nu$ and $\int_{1.97}^{1.995} \! \frac{dS}{dN} \, \mathrm{d}\nu$ (see Fig.~\ref{fig:s3}). We then defined the larger of these two integrals as the integration offset error (see Fig.\ref{fig:s3}). Second, there is an additional contribution to the error from assuming particle-hole symmetry in the calibration of $\mu$ and $\nu$. However, as shown in Fig.~\ref{fig:s3}, the particle-hole asymmetry of the entropy, $S(\nu)-S(-\nu)$, remains smaller than the integration offset error. Thus, we used the integration offset error for the error bars in Fig.~\ref{fig:1}d.
\\

For the error bars for $\mathrm{d}S/\mathrm{d}N$ in Fig.~\ref{fig:experiment_WC_entropy_integer}, we calculate the particle-hole asymmetry $|\mathrm{d}S/\mathrm{d}N(\nu) - \mathrm{d}S/\mathrm{d}N(-\nu)|$. This $\mathrm{d}S/\mathrm{d}N$ asymmetry represents errors introduced by assuming particle-hole symmetry when calibrating $\mu$ curves and determining $\mathrm{d}S/\mathrm{d}N$ via the Maxwell relation (see Fig.~\ref{fig:devPH}).
\\

\textbf{Inter-anyon distances at the gap edge}

The maximal inter-anyon distance is obtained experimentally from the filling factor value $\rm \nu^*$ of the chemical potential maxima near the incompressible state. From this filling factor value the inter-anyon distance $a$ is obtained from $\rm a=\sqrt{\frac{4\pi}{\sqrt{3}q\nu^*}}l_b$ and listed in Table I for the different fractional quantum Hall states. These inter-anyon distances are larger than the gate distance $\rm d=63$ nm such that the inter-anyon Coulomb interaction is greatly reduced in that regime.

\begin{table}[h!]
\centering
\begin{tabular}{||c c c c||}
 \hline
 $\rm \nu$ & $\rm \nu^* . 10^{-3}$ & a (nm) & a/d \\ [0.5ex]
 \hline\hline
  3/11 & 1.5 & 140 & 2.3 \\
 2/7 & 2.1 & 150 & 2.4 \\
  1/3 & 3.3 & 190 & 3.0 \\
 2/5 & 3.0 & 150 & 2.4 \\
 3/7 & 2.6 & 140 & 2.2 \\
 4/9 & 2.2 & 130 & 2.1 \\
 5/11 & 1.7 & 140 & 2.2 \\
 2/3 & 4.1 & 170 & 2.7 \\
 3/5 & 3.2 & 150 & 2.3 \\
 4/7 & 2.6 & 140 & 2.2 \\
 5/9 & 2.2 & 130 & 2.1 \\
 6/11 & 1.7 & 140 & 2.2 \\
   -1+8/17 & 2.1 & 100 & 1.6 \\
 -1+1/2 & 2.6 & 180 & 2.9 \\
  -1+7/13 & 2.0 & 110 & 1.8 \\
 -1+2/3 & 2.9 & 200 & 3.2 \\
 \hline
\end{tabular}
\caption{Inter-anyon distance near the chemical potential jump.}
\label{table:2}
\end{table}

\section{Details of theoretical calculations of entropies}

In this section, we provide details of the theoretical calculations of entropies at the half and integer fillings.

\subsection{Quantum Hall ferromagnetic states}

The zLL in the monolayer graphene has four degrees of freedom, two valleys and two spins, which will be denoted by the Pauli matrices $\tau$ and $\sigma$ respectively.
We pack the four-flavor electron operators into a four-component spinor $\Psi(\bm{r})$.
The energy scales of the interactions have a special hierarchy:
\begin{itemize}
\item The long-range part of the Coulomb interaction, parameterized by $E_C = e^2/(4\pi \varepsilon \ell_B)$, is the largest energy scale
\begin{equation}
    H_C = \frac{1}{2} \sum_{\bm{q}} V(\bm{q}) \rho(\bm{q}) \rho(-\bm{q})\,,\quad \rho(\bm{r}) = \Psi^\dag(\bm{r}) \Psi(\bm{r})\,.
\end{equation}
Note that $H_C$ is fully symmetric in the valley and spin basis and therefore the system has an $SU(4)$ symmetry at the largest energy scale.

\item The next energy scale is the anisotropy of the Coulomb interaction at short length scales. It is no longer valley symmetric and takes the following form
\begin{equation}
    H_{\text{SB}} = \frac{2\pi\ell_B^2}{2} \int d^2 r\, \sum_{\mu = x,y,z}  u_\mu \rho_\mu^2(\bm{r})\,,\quad
    \rho_\mu(\bm{r}) = \Psi^\dag(\bm{r}) (\tau_\mu \otimes \bbI) \Psi(\bm{r})\,,
    \label{eq:HSB}
\end{equation}
where $\rho_\mu(\bm{r})$ is the zLL-projected pseudo-spin density.
That \eqref{eq:HSB} is formally a contact interaction is consistent with the short-range nature of the anisotropy.
The coupling constants are $u_\mu = \frac{a}{\ell_B} E_C g_\mu$, where $a$ is the atomic lattice spcaing and $g_\mu$ some $B$-independent constants.
Through our discussion we will assume $u_x = u_{y} = u_{xy}$.
\end{itemize}
Within the zLL, the kinetic energy is completely quenched and electrons tend to anti-symmetrize their orbital wave functions to minimize the repulsion energy and thus symmetrize their valley and spin degrees of freedom.
Namely, the ground state is ferromagnetic and breaks the approximate $SU(4)$ symmetry.
The precise $SU(4)$ polarization for given experimental conditions depends on the specific form of the anisotropic interaction as well as the sublattice potentials, spin Zeeman effect and disorder of the sample.
As a result, there are either gapless Goldstone modes or soft modes with small energy gaps.
In this section, we outline a general strategy of calculating the dispersion relations of these low-energy excitations under the Hartree-Fock approximation.

The entire system is rotationally invariant and so are the dispersion relations.
Without loss of generality, we put the system on a $L_x\times L_y$ cylinder and assume the state is translationally invariant along the periodic $y$-direction.
Let $\Psi_{k,a}$ denote the fermion operator of the $k$-th zLL orbital with the flavor $a$, and we introduce the fermion two-point function
\begin{equation}
    P_{ab}(k) = \braket{\Psi_{k,a}^\dag \Psi_{k,b}}\,.
\end{equation}
In the Landau gauge, the momentum in the $y$-direction is locked with the position of the zLL orbital along the $x$-direction, and thus $P_{ab}(k)$ can be regarded as the real-space charge fluctuation.
Introducing the charge fluctuation in the momentum space $\tilde{P}(q_x) = \sum_{k} e^{-i k q_x \ell_B^2} P (k)$, we can write the Hartree-Fock energy as
\begin{equation}
\begin{gathered}
    E_{\text{HF}}[P] = \frac{1}{2A} \sum_{a,b} \sum_{q_x} \tilde{V}_{abcd}^{\text{H}}(q_x) \tilde{P}_{ad}(q_x) \tilde{P}_{bc}(-q_x) - \tilde{V}_{abcd}^{\text{F}}(q_x) \tilde{P}_{ac}(q_x) \tilde{P}_{bd}(-q_x) \\
    \tilde{V}^{\text{H}}_{abcd}(\q) = \tilde{V}_{abcd}(\q) F(\q) F(-\q)\,,\quad \tilde{V}^{\text{F}}_{abcd}(\q) = \frac{1}{N_\phi} \sum_{\q'} \tilde{V}_{abcd}(\q') F(\q') F(-\q') e^{i(q_x' q_y - q_y' q_x)\ell_B^2}
\end{gathered}
\label{eq:full HHF}
\end{equation}
where $V_{abcd}$ includes both the $SU(4)$ symmetric Coulomb interaction and the anisotropic short-range interaction.
We expect that the ground state is uniform in space, $P(k) = P$, and the interaction energy of the ground state is
\begin{equation}
\begin{gathered}
    E_{\text{HF}}^{(0)}[P] = \frac{1}{2} \sum_{a,b,c,d} \tilde{V}^H_{abcd} P_{ad} P_{bc} - \tilde{V}^F_{abcd} P_{ac} P_{bd} \\
    V^{\text{H}}_{abcd} = \frac{\tilde{V}_{abcd}(0)}{2\pi \ell_B^2}\,,\quad
    V^{\text{F}}_{abcd} = \frac{1}{N_\phi\times 2\pi \ell_B^2} \sum_{\bm{q}} \tilde{V}_{abcd}(\bm{q}) F(\bm{q}) F(-\bm{q})
\end{gathered}
\end{equation}
The ground state is dictated by minimizing $E_{\text{HF}}^{(0)}[P]$ with the appropriate sublattice potential and Zeeman energy.
Our experiment has $\ell_B = 7.55$nm and accordingly $E_c = 46.8$meV. The $SU(4)$ symmetric part of the interaction is chosen as the gate-screened Coulomb interaction projected to the zLL including corrections from higher Landau levels.
For the anisotropic part, we use $g_{xy} = -1.5, g_z = 1$ that share the same orders of magnitude with previous studies.
The sublattice and Zeeman energies are $\Delta_{AB} = 6.9$meV and  $\Delta_Z = 1.32$meV respectively, which leads to a valley and spin polarized ground state in the absence of anistropic interaction.
For $\nu=0$, including the anisotropic interaction yields a ``partially sublattice polarized" (IVC/Kekule) state and is essential for explaining the entropies at low temperatures.

The spatial fluctuation of the spectral projector $P_{ab}$ around its ground state value gives rise to the Goldstone modes or soft modes. To properly obtain the dispersion relation, we have to consider the time-dependent Hartree-Fock approximation, or equivalently, deal with the Hartree-Fock approximation of the thermal partition function
\begin{equation}
    Z = \Tr e^{-\beta H} \approx \int D[P] e^{-S_E[P]}\,,\quad S_E[P] = \int_0^{\beta} d\tau \big( i \frac{d\Gamma}{d\tau} + E_{\text{HF}}[P(\tau)] \big)
\end{equation}
where $\Gamma$ is the Berry's phase that comes from the wave function overlap $e^{-i d \Gamma} = \braket{P(\tau+ d\tau) | P(\tau)}$.
Within the Hartree-Fock approximation, orbitals are in a product state so that the calculation of the right-hand side is decomposed into calculation of each individual orbital separately.
For filling factor $\nu = -1,0$, the spectral projector has rank $(\nu+2)$ and can be written in terms of $(\nu+2)$ orthonormal four-component spinor $z_i$'s as $P = \sum_{i=1}^{\nu+2} z_i z_i^\dag$.
The Berry's phase reads
\begin{equation}
    e^{-i d \Gamma} = \exp\big( - d\tau \sum_{i=1}^{\nu+2} \sum_k z_{k,i}^\dag \frac{dz_{k,i}}{d\tau} + \calO(d\tau^2) \big)\,.
    \label{eq:general Berry phase}
\end{equation}
It is convenient to write the spectral projector in terms of $SU(4)$ rotations of its ground state value. Specifically, let $\pi_\mu$ and $\Gamma_\mu$ denote the rotation angle and generator, we have
\begin{equation}
    P(k) = e^{i \pi_\mu(k) \Gamma_\mu} P e^{-i \pi_\mu(k) \Gamma_\mu}
\end{equation}
Here the choice of $\Gamma_\mu$ depends on the ground state.
To capture low-energy excitations, it suffices to assume small rotation angle $\pi_\mu \ll 1$ and expand the action to the quadratic order. The Berry's phase becomes
\begin{equation}
    i\frac{d\Gamma}{d\tau} \approx \Tr( P \Gamma_\nu \Gamma_\mu) \pi_\mu \frac{d \pi_\nu}{d\tau}
\end{equation}
For the Hartree-Fock energy, the leading term is the ground state energy and the next term vanishes. To obtain the quadratic term, let us first expand the spectral projector
\begin{equation}
\begin{gathered}
     P(k) \approx P + \pi_\mu(k) A_\mu + \pi_\mu(k) \pi_\nu(k) B_{\mu\nu} \\
     A_\mu = i [\Gamma_\mu, P]\,,\quad B_{\mu\nu} = \Gamma_\mu P \Gamma_\nu - \frac{1}{2}  \{ \Gamma_\mu \Gamma_\nu, P \}
\end{gathered}
\end{equation}
where $A_\mu$ and $B_{\mu\nu}$ are Hermitian matrices determined by the ground state.
We define the Fourier transformation
\begin{equation}
    \tilde{\pi}_\mu(q_x) = \sum_k e^{-ik q_x \ell_B^2} \pi_\mu(k) \,,\quad
    \pi_\mu(k) = \frac{1}{N_\phi} \sum_{q_x} e^{i q_x k \ell_B^2} \tilde{\pi}_\mu(q_x)
\end{equation}
and have the quadratic order of the Hartree-Fock energy as
\begin{equation}
\begin{aligned}
    E_{\text{HF}}^{(2)} =& \frac{1}{2N_\phi} \sum_{a,b,c,d,\mu,\nu} \sum_{q_x} \Big(V^H_{abcd} (P_{ad} B_{\mu\nu, bc} + B_{\mu\nu,ad} P_{bc}) - V^F_{abcd} (P_{ac} B_{\mu\nu, bd} + B_{\mu\nu,ac} P_{bd}) \Big) \tilde{\pi}_\mu(q_x) \tilde{\pi}_\nu(-q_x) \\
    &\qquad + \Big(\frac{\tilde{V}^H_{abcd}(q_x)}{2\pi\ell_B^2} A_{\mu,ad} A_{\nu,bc} - \frac{\tilde{V}^F_{abcd}(q_x)}{2\pi\ell_B^2} A_{\mu,ac} A_{\nu,bd} \Big) \tilde{\pi}_\mu(q_x) \tilde{\pi}_\nu(-q_x)
\end{aligned}
\end{equation}
The first line is the mass term of the soft modes.
A genuine symmetry breaking ground state implies a positive-definite mass term and thus guarantees the stability of the system.
After the expansion, the Euclidean action becomes a quadratic action of the rotation angle $\pi_\mu$. We can then read out the dispersion relations from the poles in the Green's function of $\pi_\mu$.

The low-energy excitations, Goldstone modes and soft modes, are essentially particle-hole excitations of the underlying electrons and are not completely independent from each other. Therefore including interactions is important at high temperatures and necessary for explaining the saturation of entropies.
In our calculation of the entropy, we ignore the interactions for the sake of simplicity, which is a reasonable approximation at low enough temperature and starts to break down when the number of thermally excited bosons per flux is of the order of unit. Fig.~\ref{fig:supp theory} shows that the approximation is more accurate for temperatures below $60K$.

\subsection{Non-interacting composite fermion gas}

To analyze the $\mathrm{d}S/\mathrm{d}N$ jumps observed in the fractional quantum Hall states, we compared the data with the entropy of a non-interacting composite fermion system. The blue curve in Fig.3 b were obtained using a non-interacting density of states $\rho(E,B_*)$, which evolves from a Heaviside step function-like shape at zero $B_*$ to quantized levels ($\lambda$-levels) as $B_*$ increases. The density of states is given by:
\begin{equation}
\rho(E,B_*) = \frac{eB_*}{h} \sum_{N_L=0}^{\infty} g(E,(N_L+1/2)\hbar \omega^{*}_{c},\Gamma),
\label{eq:cf_dos}
\end{equation}
where $\hbar \omega^{*}_{c} = eB_*/m^{*}$ is composite fermion cyclotron energy and $\frac{eB_*}{h}$ is the degeneracy of $\lambda$-levels. In Eq.\eqref{eq:cf_dos}, the Gaussian function $g(E,(N_L+1/2)\hbar \omega^{*}_{c},\Gamma) = \frac{1}{\Gamma\sqrt{2\pi}} e^{-\frac{(E-(N_L+1/2)\hbar \omega_{c})^{2}}{2\Gamma^{2}}}$ represents the $N_L$th $\lambda$-levels with disorder broadening characterized by $\Gamma$. In our model, we used the composite fermion effective mass $m^*=0.25m_e$ and disorder broadening $\Gamma=155\mu eV$, estimated from the FQH energy gaps measured in Fig.3a. The entropy is then calculated as:
\begin{equation}
S(n,B_*)=k_B\int \rho(E,B_*) \left [f_{FD}(E)
\ln(f_{FD}(E))+(1-f_{FD}(E))ln(1-f_{FD}(E)) \right] dE,
\label{eq:cf_S_2d}
\end{equation}
where $f_{FD}(E) = 1/(1+e^{\frac{E-\mu}{k_{B}T}})$ is the Fermi-Dirac distribution. In Eq.\eqref{eq:cf_S_2d}, the dependence of the entropy on electron density $n$ is determined by varying the chemical potential $\mu$. The electron density $n$ is given by:
\begin{equation}
n=\int \rho(E,B_*) \times f_{FD}(E) dE.
\label{eq:cf_density}
\end{equation}

Solving Eqs.\eqref{eq:cf_S_2d} and \eqref{eq:cf_density} gives the entropy as a function of $n$ and $B_*$. In the experiment, a constant perpendicular magnetic field $B_\perp$ was applied, making the effective magnetic field $B_*$ experienced by composite fermions dependent on the density $n$, as described by the equation below:
\begin{equation}
B_{\perp} = B_*-2\phi_0 n,
\label{eq:flux_attachment}
\end{equation}
where $2\phi_0$ represents two flux quanta attached to each composite fermion. By inserting the flux attachment condition in Eq.\eqref{eq:flux_attachment} into Eq.\eqref{eq:cf_S_2d}, we derive the entropy as a function the density. Fig.~\ref{fig:freeCFm}a shows composite fermion entropy $S$ calculated as a function of $\nu$ at $B_\perp$ = 13.8 T and for different disorder broadening $\Gamma$. The composite fermion entropy exhibits dome structures reminiscent to the integer quantum Hall entropy shown in Fig. 1c. To derive $\mathrm{d}S/\mathrm{d}N$, we compute the derivative of $S$ (see Fig.~\ref{fig:freeCFm}b). $\mathrm{d}S/\mathrm{d}N$ shows jumps at fractional $\nu$, consistent with the dome structures observed in $S$. As shown in Fig.~\ref{fig:freeCFm}a and b, the entropy of the fully gapped $\nu$ = 1/3 and 2/5 states decreases with increasing $\Gamma$, as the quasi-particles near the band tail become localized when $k_{B} T$ drops below the disorder potential (approximately $\Gamma/2$). In addition, the $\mathrm{d}S/\mathrm{d}N$ jumps at weak fractional quantum Hall states, such as the $\nu$ = 5/11 and 6/13 states, also become less pronounced as the energy gaps of these states become comparable to $\Gamma$.

To identify the effect of the disorder broadening on the weak FQH states with overlapping $\lambda$-levels in energy, we compute the contribution of each $N_L$th $\lambda$-level to the total entropy as a function of $\nu$. To calculate the contribution of the various $N_L$th $\lambda$-levels to the total entropy, we used a single $N_L$th $\lambda-$-level density of states Eq.\eqref{eq:cf_S_2d}, as expressed below.
\begin{equation}
\rho^{N_L}(E,B_*) = \frac{eB_*}{h} g(E,(N_L+1/2)\hbar \omega^{*}_{c},\Gamma),
\label{eq:cf_single_dos}
\end{equation}

Figs. \ref{fig:freeCFm}c and d show the entropy contribution of each $N_L$th $\lambda$-levels to the total CF entropy as a function of $\nu$ (see the colored lines). At $\nu<1/3$ and $1/3<\nu<2/5$, the entropy of the $N_L$=0 and $N_L$=1 $\lambda$-levels fully accounts for the total entropy, respectively, because $\lambda$-levels are fully quantized at large $B_*$ which is inversely proportional to $\nu$. On the other hand, at larger $\nu$, the total entropy consists of contributions from multiple overlapping $\lambda$-levels, as illustrated by the overlaps between the different dome structures in \ref{fig:freeCFm}c. In these weak FQH states, the jumps in $\mathrm{d}S/\mathrm{d}N$ are smoothed out.

\subsection{Composite fermi liquid interacting with gauge fields}

When the monolayer graphene is at half integer fillings, such as $\nu = -3/2$ and $\nu = -1/2$, it is likely that the low-energy excitations can be captured by the theory of a half-filled Landau level.
An exact modeling of the $\nu = -1/2$ is much more complicated and is not the focus of this section.
Here we simply review the classic Halperin-Lee-Read (HLR) approach of a single-component half-filled Landau level and explain how the entropy is calculated.

The HLR theory is based on the theory of composite fermion. Let $\psi$ be the composite fermion field that is coupled to both the dynamical gauge field $a$ as well as the external gauge field $A$.
The Euclidean Lagrangian $\calL_E = \calL_0 + \calL_{int}$ has two parts, the kinetic part
\begin{equation}
	\calL_0 = \bar{\psi}(\hbar \partial_\tau - \mu + i e a_0) \psi - \frac{1}{2m_b} \bar{\psi} (\hbar \partial_i + i e a_i  + i e A_i)^2 \psi - \frac{i}{4\pi} \frac{e^2}{\hbar} a_0 \varepsilon^{ij} \partial_i a_j\,,
\end{equation}
and the screened Coulomb interaction
\begin{equation}
	\calL_{int} = \frac{1}{2} \frac{e^2}{(4\pi \hbar)^2} \int d^2 r' (\nabla\times\bm{a}(r)) V(r-r')  (\nabla\times\bm{a}(r'))\,.
\end{equation}
Note that the mass of the composite fermion $m_b$ generally differs from the electron mass.
The equations of motion of $a_0$ dictates the flux attachment rule
\begin{equation}
	\bar\psi \psi = \frac{e}{4\pi \hbar} \nabla\times \bm{a}
\end{equation}
namely each composite fermion binds itself with two magnetic flux quanta of the dynamical gauge field.

The HLR theory is based on the assumption that the system can be characterized in a perturbative approach.
The mean-field solution is determined by the following saddle-point equation
\begin{equation}
	a_0 = 0\,,\quad \bm{a} + \bm{A} = 0\,.
\end{equation}
Namely, the dynamical gauge field exactly cancels the background magnetic field and the composite fermion does not see any effective magnetic field
\begin{equation}
    \calL_{MF}[\bar\psi,\psi,a,A] = \calL_{MF}[\bar\psi,\psi] = \bar{\psi}(\hbar \partial_\tau - \mu) \psi - \frac{1}{2m_b} \bar{\psi} (\hbar \partial_i)^2 \psi
\end{equation}
The fluctuation away from this saddle point is assumed to be so small that it is legitimate to perform perturbation calculation around it.
As a result, the entropy of the half-filled Landau level comprises of two parts, that from the composite fermion and that from the dynamical gauge field. The former is formally the same as the entropy of a free non-relativistic electron gas, which yields a linear-$T$ behavior at low temperatures.
The later depends on the specific form of the interaction. Calculations using the random phase approximation show that the short-range interactions, which the screened Coulomb interaction belongs to, yield a $T^{2/3}$ non-Fermi liquid behavior at low temperatures.

\subsection{Electron Wigner crystal and quasiparticle Wigner crystal}

At filling factors close to incompressible states, the entropy per particle is significantly suppressed at low temperatures due to the formation of charge order. In this appendix, we calculate the entropy of the Wigner crystal (WC) formed by electrons or fractionally charged quasiparticles. Our framework and treatment of screening and disorder closely follow Ref.~\onlinecite{assouline_energy_2024}, originally developed for zero-temperature physics. Here, we demonstrate how the same framework can be extended to finite temperature to handle phonon excitations of the WC.

We begin by examining the electron Wigner crystal in monolayer graphene (MLG). In MLG, the cyclotron gap is much larger than the Coulomb interaction energy, which allows us to safely project the Lagrangian onto the zero-energy Landau level (zLL). We construct a trial many-body wave function for electrons arranged on a triangular lattice at positions $\mathbf{R}_i$,
\begin{equation}
    \Psi_{\{\mathbf{R}_i\}}(\{\mathbf{r}_i\}) = \hat{\mathcal{A}} \prod_i \phi_{\mathbf{R}_i}(\mathbf{r}_i),
\end{equation}
where $\phi_{\mathbf{R}_i}(\mathbf{r}_i)$ is the zLL wave function centered at $\mathbf{R}_i$, and $\hat{\mathcal{A}}$ denotes antisymmetrization. Using this wave function, we evaluate the Lagrangian
\begin{align}
L_{\mathrm{WC}} &= \frac{i \hbar}{2}
\frac{\langle \Psi \mid \dot{\Psi} \rangle - \langle \dot{\Psi} \mid \Psi \rangle}
     {\langle \Psi \mid \Psi \rangle}
- \frac{\langle \Psi \mid \hat{V} \mid \Psi \rangle}{\langle \Psi \mid \Psi \rangle}\\[6pt]
&= -E_{\mathrm{WC}}
- \frac{\hbar}{2 \ell_B^2} \sum_i (\mathbf{u}_i \times \hat{z}) \cdot \dot{\mathbf{u}}_i
+ \frac{1}{4} \sum_{i \neq j} \frac{\partial^2 V_H(|\mathbf{R}_i - \mathbf{R}_j|)}{\partial R^\alpha \partial R^\beta}
\,(u_i^\alpha - u_j^\alpha)\,(u_i^\beta - u_j^\beta),
\end{align}
where $E_{\mathrm{WC}} = \frac{N}{2} \sum_{\mathbf{R}_i \neq 0} V_H(|\mathbf{R}_i|)$ is the ground-state energy of the Wigner crystal, and $N$ is the number of electrons. Phonon excitations are described by the displacements $\mathbf{u}_i = \mathbf{R}_i - \mathbf{R}_i^0$, where $\mathbf{R}_i^0$ are the equilibrium positions. We expand the Lagrangian to second order in $\mathbf{u}_i$, neglecting Fock terms in accordance with Ref.~\cite{maki_static_1983}. The Hartree potential $V_H(\mathbf{R})$ is related to the bare Coulomb potential $V(\mathbf{R})$ in Fourier space via $V_H(\mathbf{q}) = V(\mathbf{q})\bigl|F(\mathbf{q})\bigr|^4$, where $F(\mathbf{q}) = e^{-q^2 \ell_B^2 / 4}$ is the zLL form factor. The classical treatment of the WC is valid when electrons are sufficiently far apart such that exchange interactions become negligible. In practice, we focus on the filling range $\delta \nu < 0.3$, corresponding to inter-electron distances $R > 5 \ell_B$.

To determine the phonon spectrum, we transform to momentum space
\begin{equation}
\mathbf{u}_i
= \frac{1}{\sqrt{N}} \sum_{\mathbf{q} \in \mathrm{BZ}} e^{i \mathbf{q} \cdot \mathbf{R}_i^0} \mathbf{u}(\mathbf{q}),
\end{equation}
where $\mathrm{BZ}$ denotes the Brillouin zone, $\mathbf{R}_i^0$ are equilibrium lattice positions, and $N$ is the total number of particles. The Lagrangian in momentum space becomes
\begin{equation}
L_{\mathrm{WC}}
= \frac{1}{2} \sum_{\mathbf{q} \in \mathrm{BZ}}
\left( \frac{\mathbf{u}_{-\mathbf{q}}}{a_0} \right)^\mathrm{T}
\left[
  i \hbar \omega \frac{a_0^2}{\ell_B^2} i \sigma^y
  -
  \mathbf{M}_{\mathbf{q}}
\right]
\frac{\mathbf{u}_{\mathbf{q}}}{a_0},
\label{eq:LWC_qspace}
\end{equation}
where $a_0$ is the inter-particle distance determined by the filling factor $\nu$, and $\sigma^y$ is a Pauli matrix in the basis of $(u_x,u_y)^T$. The dynamical matrix $\mathbf{M}_{\mathbf{q}}$ is given by
\begin{equation}
\mathbf{M}_{\mathbf{q}}
= - \frac{1}{a_c a_0^2} \sum_{\mathbf{G}}
\bigl(\mathbf{G} a_0\bigr)v_{\mathbf{G}}
\bigl(\mathbf{G} a_0\bigr)^\mathrm{T}
+
\sum_{\mathbf{G}}
\bigl(\mathbf{q}a_0 + \mathbf{G}a_0\bigr)
v_{\mathbf{q} + \mathbf{G}}
\bigl(\mathbf{q}a_0 + \mathbf{G}a_0\bigr)^\mathrm{T},
\label{eq:Mq}
\end{equation}
with $a_c = \sqrt{3}/2$ being the geometric factor for a triangular lattice, $\mathbf{G}$ the reciprocal lattice vectors, and $v_{\mathbf{q}}$ the Fourier transform of $V_H(\mathbf{R})$. The phonon dispersion relations then follow from the zero modes of $\mathbf{M}_{\mathbf{q}}$:
\begin{equation}
\hbar \omega_{\mathbf{q}}
= \frac{\ell_B^2}{a_0^2}
\sqrt{M_{\mathbf{q}}^{00} M_{\mathbf{q}}^{11}-\bigl(M_{\mathbf{q}}^{01}\bigr)^2}.
\label{eq:phonon_dispersion}
\end{equation}

The free energy of the Wigner crystal is
\begin{equation}
\frac{F_{\mathrm{WC}}}{N}
= \frac{E_{\mathrm{WC}}}{N}
+ \frac{1}{N} \sum_{\mathbf{q} \in \mathrm{BZ}}
\left[
  \frac{1}{2}\,\hbar \omega_{\mathbf{q}}
  + k_B T \ln \left( 1 - e^{-\hbar \omega_{\mathbf{q}} / (k_B T)} \right)
\right],
\label{eq:FWC}
\end{equation}
where $k_B$ is the Boltzmann constant, $T$ is temperature, and $\omega_{\mathbf{q}}$ are the phonon frequencies obtained from the dynamical matrix. From the free energy, we compute the entropy per particle
\begin{equation}
\frac{S_{\mathrm{WC}}}{N}
= - \frac{k_B}{N}\sum_{\mathbf{q} \in \mathrm{BZ}}
\left[
  \ln \Bigl(1 - e^{-\hbar \omega_{\mathbf{q}}/(k_B T)}\Bigr)
  +
  \frac{\hbar \omega_{\mathbf{q}}/(k_B T)}{e^{\hbar \omega_{\mathbf{q}}/(k_B T)} - 1}
\right].
\label{eq:SWC}
\end{equation}
and the chemical potential
\begin{equation}
\mu_{\mathrm{WC}}^{\mathrm{int}}
= \frac{1}{\partial N} \left(F_{\mathrm{WC}}
  - \frac{e^2N^2}{2C_g}\right),
\label{eq:muWC}
\end{equation}
where $e^2N^2/2C_g$ is the classical charging energy. We subtract this term in the definition of the chemical potential following the experimental convention.

Although we do not explicitly consider excitations involving higher Landau levels (LLs), we account for their screening effects on the interaction. The bare Coulomb interaction is screened by the encapsulating hexagonal boron nitride (hBN), the graphite gates, and inter-LL transitions. We write the screened Coulomb potential as
\begin{equation}
\label{eq:V0q}
V_0(\mathbf{q}) = E_C \ell_B \frac{4\pi \sinh(\beta d_t q)\,\sinh(\beta d_b q)}{\sinh[\beta(d_t + d_b)q]\,q},
\end{equation}
where $E_C = e^2/4\pi \epsilon_0 \epsilon_{\mathrm{hBN}} \ell_B$ is the Coulomb energy scale, and $d_t = d_b = \SI{60}{nm}$ are the distances from the graphene layer to the top and bottom graphite gates. We adopt dielectric constants for hBN of $\epsilon_{\perp} = 6.6$ (out-of-plane) and $\epsilon_{\parallel} = 3.15$ (in-plane), giving $\beta = \sqrt{\epsilon_{\parallel}/\epsilon_{\perp}}$ and $\epsilon_{\mathrm{hBN}} = \sqrt{\epsilon_{\perp}\,\epsilon_{\parallel}}$. The static dielectric response from inter-LL virtual excitations is obtained within the random phase approximation (RPA)
\begin{equation}
\label{eq:Vscreened}
V(\mathbf{q}) = \frac{V_0(\mathbf{q})}{1 - V_0(\mathbf{q}) \Pi_\nu(\mathbf{q}, \omega=0)},
\end{equation}
where $\Pi_\nu(\mathbf{q})$ is the polarizability at filling factor $\nu$, computed following Refs.~\cite{shizuya_electromagnetic_2007, misumi_electromagnetic_2008, gorbar_dynamics_2010, papic_topological_2014, zibrov_tunable_2017}.

To compute the entropy of Wigner crystals formed by fractionally charged quasiparticles, several modifications are needed. Within the composite fermion picture, quasiholes in fractional states such as $\nu = 1/3$ have wave functions identical to those in the lowest Landau level, except with $\ell_B$ replaced by $\tilde{\ell}_B = \sqrt{\hbar/e^* B}$. In the filling range of interest, the quasiparticles are sufficiently far apart that the classical treatment remains valid, so we estimate the entropy near fractional quantum Hall states by substituting $e \to e^*$ and $\ell_B \to \tilde{\ell}_B$. Note that $\tilde{\ell}_B$ not only appears in the form factor $F(\mathbf{q})$ but also modifies the guiding-center commutation relation $[X, Y] = i \tilde{\ell}_B^2$ and therefore the flux counting.

If these quasiparticles exhibit non-Abelian statistics, there is an additional topological contribution to the entropy
\begin{equation}
\label{eq:Stopo}
S_{\mathrm{total}} = S + S_{\mathrm{topo}},
\quad
S_{\mathrm{topo}} = N_a k_B \ln(\sqrt{\mathcal{D}}),
\end{equation}
where $N_a$ is the number of non-Abelian quasiparticles and $\mathcal{D}$ is their quantum dimension. A principal experimental goal is to detect this non-Abelian entropy $S_{\mathrm{topo}}$.

An additional complication arises from screening due to composite fermion $\Lambda$ levels, whose energies are comparable to the Coulomb interaction. To capture this effect, we employ a phenomenological model for the RPA polarizability $\Pi_\nu(\mathbf{q})$:
\begin{equation}
\label{eq:pheno}
\Pi_{\mathrm{FQH}}^\chi(\mathbf{q}) = \frac{4 \ln 4}{2 \pi} \tanh\bigl(\chi q^2 \ell_B^2\bigr) \frac{1}{\hbar \omega_c},
\end{equation}
where the parameter $\chi$ controls the strength of screening and is treated as a fitting parameter for the quasiparticle Wigner crystal, and $\hbar \omega_c$ denotes the cyclotron gap. We find that $\chi = 0.62$ reproduces $\Pi_{\mathrm{LL}}(q)$ for bilayer graphene. Moreover, the screening parameter $b$ in the main text can be related to $\chi$ by expanding $\Pi_{\mathrm{FQH}}^\chi(\mathbf{q})$ in powers of $q$, yielding $\Pi_{\mathrm{FQH}}(q) = -b q^2 + \cdots$. This approach reproduces the experimentally measured chemical potential reasonably well~\cite{yang_experimental_2021}.

Finally, the chemical potential smooths out near incompressible states, suggesting that disorder significantly affects the chemical potential and, consequently, the entropy of the Wigner crystal. Disorder can play two roles: it can induce a slowly varying chemical potential, resulting in a disorder-averaged chemical potential in capacitive measurements; and it can gap out the phonon excitations, leading to a strong suppression of entropy at low temperatures. The latter effect is difficult to capture for generic interactions~\cite{giamarchi_disordered_2002} and requires future experiments to measure the phonon gap directly. The first effect, however, can be described using a slowly varying disorder model developed in Ref.~\onlinecite{assouline_energy_2024} within the local density approximation (LDA). Specifically, the disorder-averaged chemical potential $\bar{\mu}_{\mathrm{int}}$ is given by:
\begin{equation}
\bar{\mu}^{\mathrm{int}}(V_g) = \int dV_D\, P(V_D)\, \mu^{\mathrm{int}}_{\mathrm{FQH}}\bigl(n(V_D)\bigr),
\quad
\mu^{\mathrm{int}}_{\mathrm{FQH}}(\nu)=
\begin{cases}
\frac{1}{2}\Delta + \mu_{\mathrm{WC}}^{\mathrm{int}}\bigl(\nu-\nu_0\bigr), & \nu>\nu_0, \\
-\frac{1}{2}\Delta + \mu_{\mathrm{WC}}^{\mathrm{int}}\bigl(\nu-\nu_0\bigr), & \nu<\nu_0,
\end{cases}
\end{equation}
where $V_g$ is the gate voltage, $\mu_{\mathrm{int}}^{\mathrm{FQH}}(\nu)$ is the clean-limit chemical potential incorporating the fractional quantum Hall gap $\Delta$, $n(V_D)$ is the charge density in the presence of an external potential $V_D$, and $P(V_D)$ is the disorder distribution. In practice, we assume a Gaussian disorder distribution $P\bigl(V_D\bigr)=\frac{1}{\sqrt{2\pi}\,\Gamma}\exp\bigl(-V_D^2/(2\Gamma^2)\bigr)$ and treat both the gap $\Delta$ and the broadening $\Gamma$ as fitting parameters.

This inhomogeneous broadening effect also complicates the extraction of $S_{\mathrm{topo}}$ and $\mathcal{D}$. When quasiparticles and quasiholes coexist in the system due to disorder, the number of electrons added, $\Delta N$, no longer directly measures the number of non-Abelian excitations, $N_a$. In the case of the $\nu_0 = -1 + 1/2$ state,
\begin{equation}
\Delta N = 4(N_+ - N_-), \quad N_a = N_+ + N_-,
\end{equation}
where $N_+$ and $N_-$ are the number of quasiparticles and quasiholes, respectively. We can quantify this effect by computing $N_a(N)$ using the broadening $\Gamma$ obtained from fitting the experimental data. As shown in Fig.~\ref{fig:Na}, $\mathrm{d}N_a = 4\,\mathrm{d}N$ remains valid down to $\delta \nu = 0.002$, which provides a substantial range of $\delta \nu$ where $\mathrm{d} S_{\mathrm{topo}}/ \mathrm{d} N$ is unaffected by inhomogeneous broadening.

We next review the fitting procedure and the resultant parameters $\Delta$, $\Gamma$, and $\chi$ (the screening parameter), as presented in Ref.~\cite{assouline_energy_2024}. We choose to fit the chemical potential $\mu(\nu)$ at low temperature, where experimental measurements are most reliable and the lineshape of $\mu(\nu)$ is less sensitive to small variations in the parameters. The final values of $\Delta$, $\Gamma$, and $\chi$ are shown in Table~\ref{tab:parameter}.

\begin{table}[h!]
\renewcommand*{\arraystretch}{1.3}
\newcolumntype{C}{>{\centering\arraybackslash}X}
\centering
\begin{tabularx}{0.55\textwidth}{C|CCC}
\hline\hline
 & $\Delta$ (meV) & $\Gamma$ (meV) & $\chi$ \\
\hline
$\nu=-1+1/2$ & 2.4 & 1.0 & 0.85 \\
$\nu=-1+2/3$ & 3.0 & 1.2 & 0.92 \\
$\nu=2/5$ & 3.7 & 1.9 & 0.17 \\
\hline\hline
\end{tabularx}
\caption{Fitting parameters of the disordered Wigner crystal model for fractional fillings. $\Delta$ is the thermodynamic gap, $\Gamma$ is the disorder broadening, and $\chi$ is the effective screening strength defined in Eq.~(\ref{eq:pheno}).}
\label{tab:parameter}
\end{table}

After obtaining these parameters, we show
\begin{equation}
\frac{d S}{d N} \approx \frac{\bar{\mu}_{\mathrm{int}}(V_g, T_{\mathrm{hot}}) - \bar{\mu}_{\mathrm{int}}(V_g, T_{\mathrm{cold}})}{\Delta T}
\end{equation}
in Fig.~4 of the main text to enable a more quantitative comparison with experiment.

\clearpage

\section{Supplementary Figures}

\begin{figure*}[ht]
    \centering
    \includegraphics[width = \textwidth]{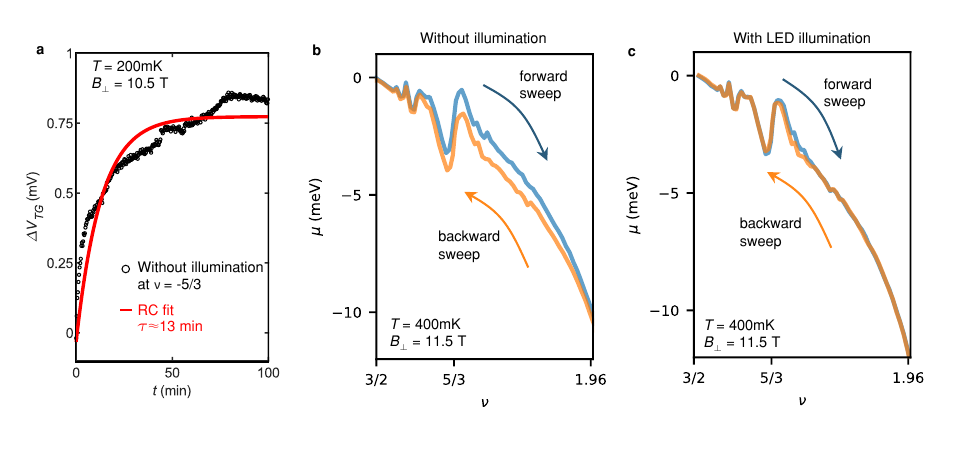}
    \caption{\textbf{Slow charging rate and the effect of LED illumination in the monolayer graphene sample}. \textbf{a} The change of the top gate $\Delta V_{TG}$ at the conductance minimum of the detector layer over time $t$. The sample layer is fixed at $\nu=-5/3$. The red line is a RC charging curve fit: $\Delta V_{TG} \propto (1-e^{-t/{\tau}})$, where the time constant $\tau$ is 13 min.  \textbf{b} Hysteresis in $\mu$ when the chemical potential measurement was taken without illumination. The colored arrows mark the sweep direction of the sample layer density $\nu$. \textbf{c} When the same chemical potential measurement is taken under LED illumination, hysteresis in $\mu$ disappears.
    }
    \label{fig:s1}
\end{figure*}

\begin{figure*}[ht]
    \centering
    \includegraphics
    {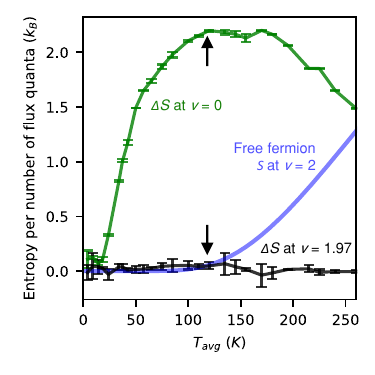}
    \caption{\textbf{The relative change of entropy $\Delta S$ at high temperatures}. The black and green lines are $\Delta S$ at $\nu=-1.97$ and 0 measured as a function of $T_{avg}$. The decrease in $\Delta S(\nu=0)$ at a high temperature is attributed to an increase in $S(\nu\approx-2)$ (see Method for details). The blue line represents $S(\nu=-2)$ calculated from the free particle model. The cyclotron energy $\hbar\omega_{c}$=1688K and the disorder broadening $\Gamma$=150 $\mu$eV are assumed. The free particle model shows that $S(\nu=-2)$ remains zero at $T_{avg} \lesssim $ 120K and rapidly increases at a higher temperature.
    }
    \label{fig:s2}
\end{figure*}

\begin{figure*}[ht]
    \centering
    \includegraphics[width = 170mm]{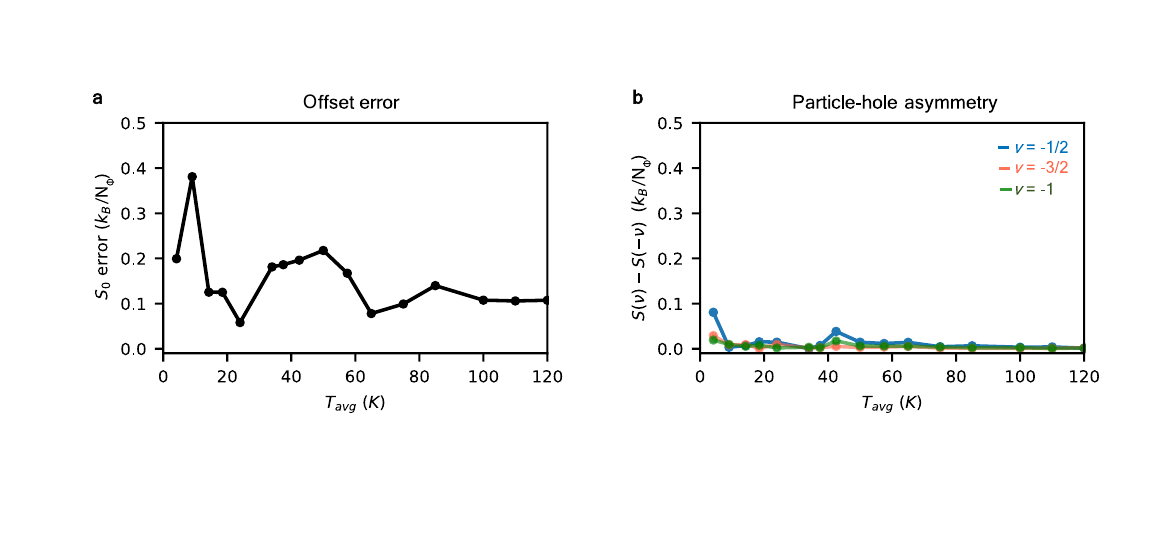}
    \caption{\textbf{Error bars estimation for $\Delta S$}. There are two sources of error in our entropy data, $\Delta S = S - S_0$. The first is the error in the integration offset $S_0$ which arises from referencing the entropy at $\nu = 1.97$ rather than at $|\nu| = 2$. The second source is the error in $S$ caused by the particle-hole symmetry assumption used in our calibration. \textbf{a}, The offset error as a function $T_{avg}$. The offset error is estimated as the larger of the two integrals: $\int_{-1.995}^{-1.97} \frac{dS}{dN} , \mathrm{d}\nu$ and $\int_{1.97}^{1.995} \frac{dS}{dN} , \mathrm{d}\nu$ (see Method for a detailed explanation). \textbf{b}, The particle-hole asymmetry $|S(\nu)-S(-\nu)|$ as a function of $T_{avg}$ at $\nu=$ -1/2, -3/2, and -1. The error bars in Fig. 1d represent the offset error shown in \textbf{a}, as this error is larger than the particle-hole asymmetry shown in \textbf{b}.
    }
    \label{fig:s3}
\end{figure*}

\begin{figure*}[ht]
    \centering
    \includegraphics[width = 170mm]{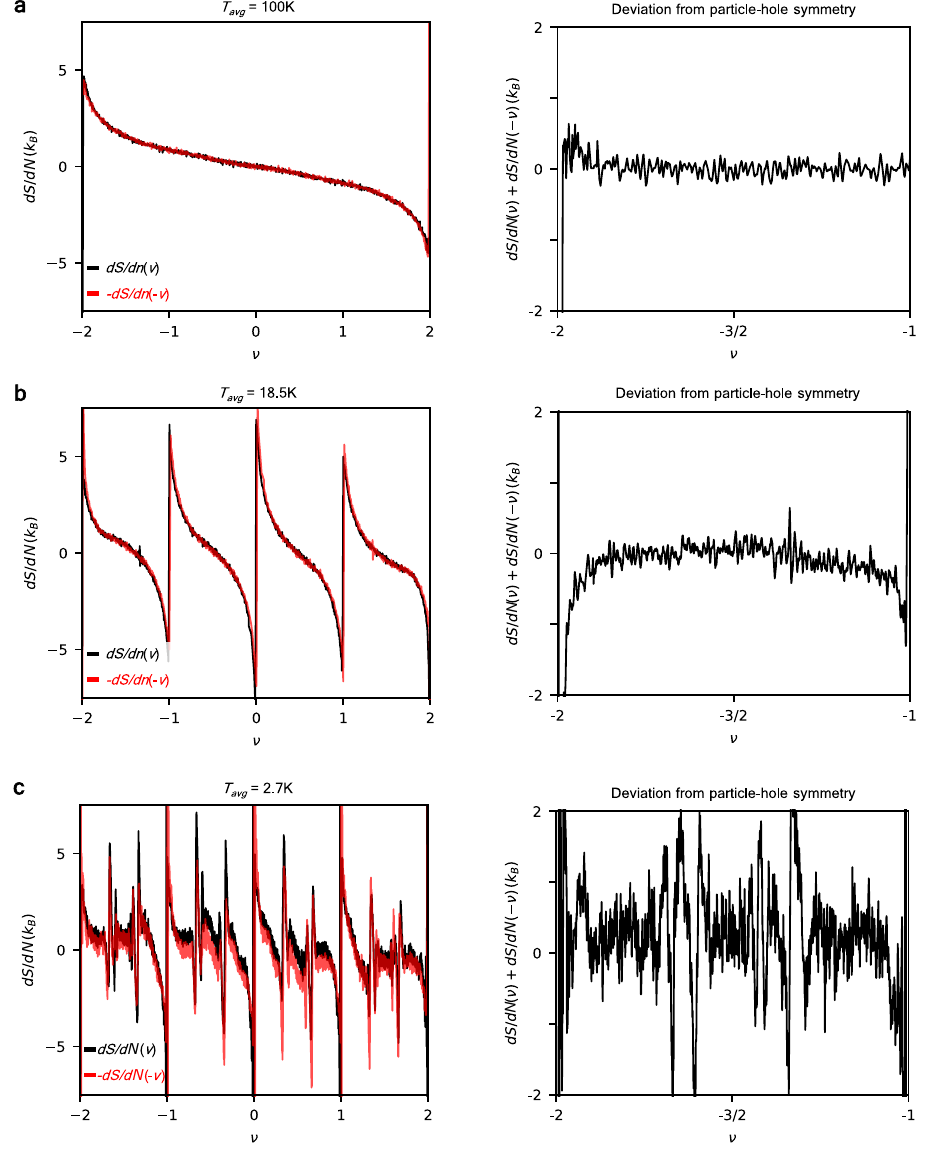}
    \caption{\textbf{Error bars estimation for $\mathrm{d}S/\mathrm{d}N$}. The left panels shows $\mathrm{d}S/\mathrm{d}N$ (black) and the particle-hole symmetrized (red) curves at $T_{avg} =$ 100K, 18.5K, and 2.7K. The right panels show the deviation from particle-hole symmetry $\mathrm{d}S/\mathrm{d}N(\nu) + \mathrm{d}S/\mathrm{d}N(-\nu)$. The particle-hole asymmetry of $\mathrm{d}S/\mathrm{d}N$ is used for the error bars in main text Fig.~\ref{fig:experiment_WC_entropy_integer}ab.
    }
    \label{fig:devPH}
\end{figure*}

\begin{figure*}[ht]
    \centering
    \includegraphics[width = 183mm]{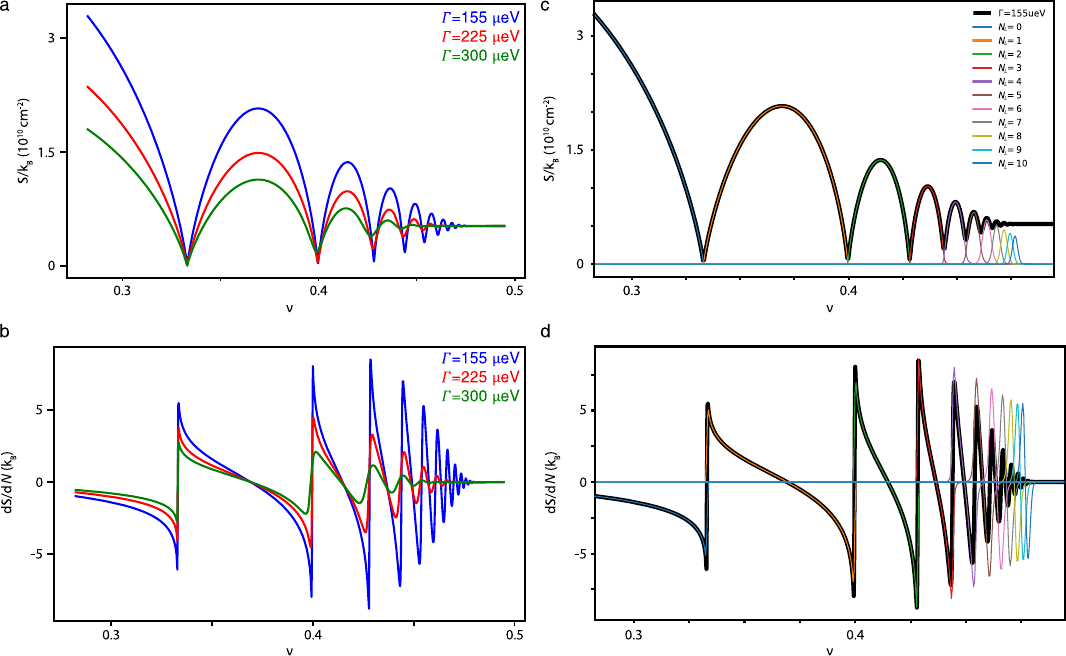}
    \caption{\textbf{Calculated $\mathrm{d}S/\mathrm{d}N$ for composite fermions.} \textbf{a}, Composite fermion entropy $S$ calculated as a function of $\nu$ at $B_\perp$ = 13.8 T and for different disorder broadening $\Gamma$. The composite fermion entropy exhibits dome structures similar to those observed in the integer quantum Hall entropy in Fig. 1. \textbf{b}, Composite fermion $\mathrm{d}S/\mathrm{d}N$ obtained by taking the derivative of $S$ in \textbf{a}. Similar to the integer quantum Hall states, $\mathrm{d}S/\mathrm{d}N$ shows jumps at fractional $\nu$. At low $\Gamma$, the $\mathrm{d}S/\mathrm{d}N$ follows a free particle logarithmic function $\ln[x/(1-x)]$ near the fully gapped states at $\nu$= 1/3 and 2/5. On the other hand, with increasing $\Gamma$, the jumps in $\mathrm{d}S/\mathrm{d}N$ become less pronounced.}
    \label{fig:freeCFm}
\end{figure*}

\begin{figure*}[ht]
    \centering
    \includegraphics[width = 170mm]{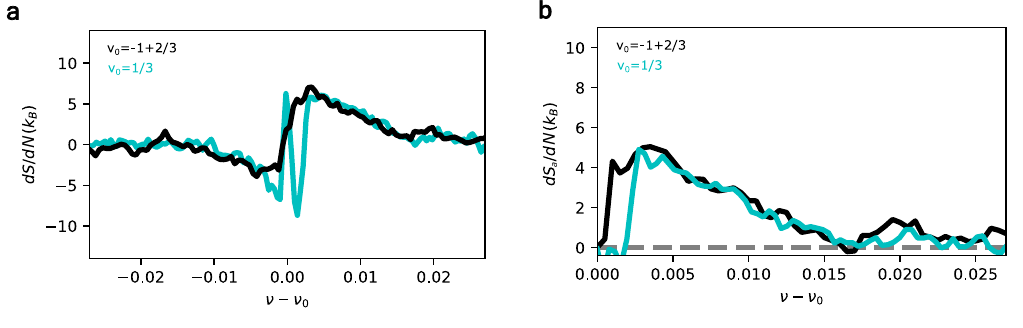}
    \caption{\textbf{Comparison of the entropy per particle at 1/3 ($\Delta \mu_{1/3}=1.9$meV) and -1+2/3 ($\Delta \mu_{-1+2/3}=1.3$meV).} \textbf{a} $\mathrm{d}S/\mathrm{d}N$ measured as a function $\nu-\nu_0$ obtained by substracting two chemical potential curves obtained at T=300mK and T=100mK. \textbf{b} Anti-symmetrized entropy per particle $dS_a/dN=(\mathrm{d}S/\mathrm{d}N(\nu)-\mathrm{d}S/\mathrm{d}N(-\nu))/2$.
    }
    \label{fig:6}
\end{figure*}

\begin{figure*}[ht]
    \centering
    \includegraphics[width = 90mm]{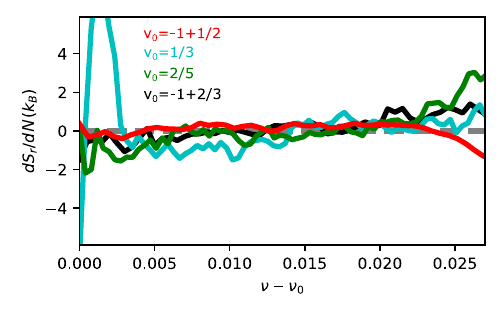}
    \caption{\textbf{Symmetrized entropy per particle from chemical potential curves obtained at 300mK and 100mK.} Defined as $dS_r/dN=(\mathrm{d}S/\mathrm{d}N(\nu-\nu_0)+\mathrm{d}S/\mathrm{d}N(-\nu-\nu_0))/2$.
    }
    \label{fig:7}
\end{figure*}

\begin{figure*}[ht]
    \centering
    \includegraphics[width = 170mm]{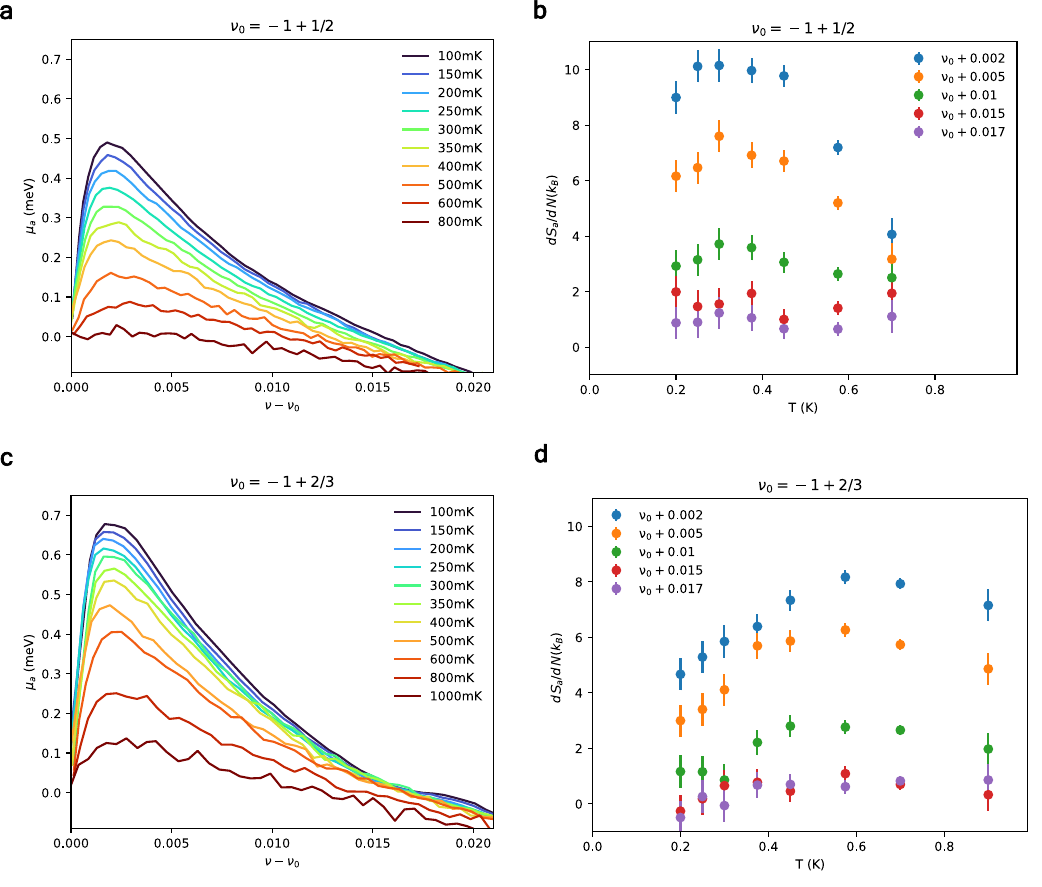}
    \caption{\textbf{Chemical potential and entropy per particle as a function of temperature}  \textbf{a} Anti-symmetrized chemical potential $ \mu_a=[\mu(\nu-\nu_0)-\mu(-\nu-\nu_0)]/2$ at different temperature near -1/2.
     \textbf{b} Anti-symmetrized entropy per particle, obtained over a temperature difference $\geq$200mK, as a function of temperature for different doping near -1/2.  \textbf{c} Anti-symmetrized chemical potential $\mu_a$ at different temperature near -1/3.
     \textbf{d} Anti-symmetrized entropy per particle, obtained over a temperature difference $\geq$200mK, as a function of temperature for different doping near -1/3.
    }
    \label{fig:8}
\end{figure*}

\begin{figure*}
\centering
\includegraphics[width=0.25\textwidth]{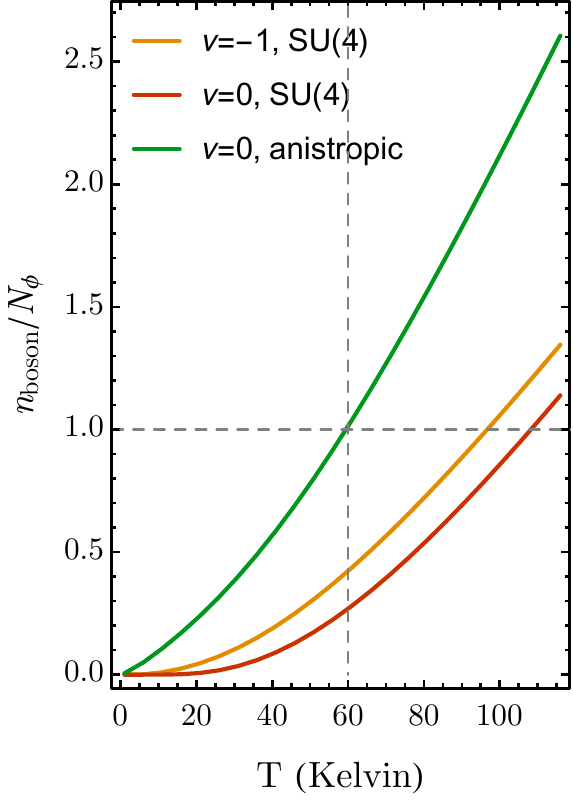}
\caption{
Number of thermally excited bosons per flux. We choose the large momentum cutoff such that there are $N_\phi$ different momenta included in the calculation.}
\label{fig:supp theory}
\end{figure*}

\begin{figure*}
\centering
\includegraphics[width=0.3\textwidth]{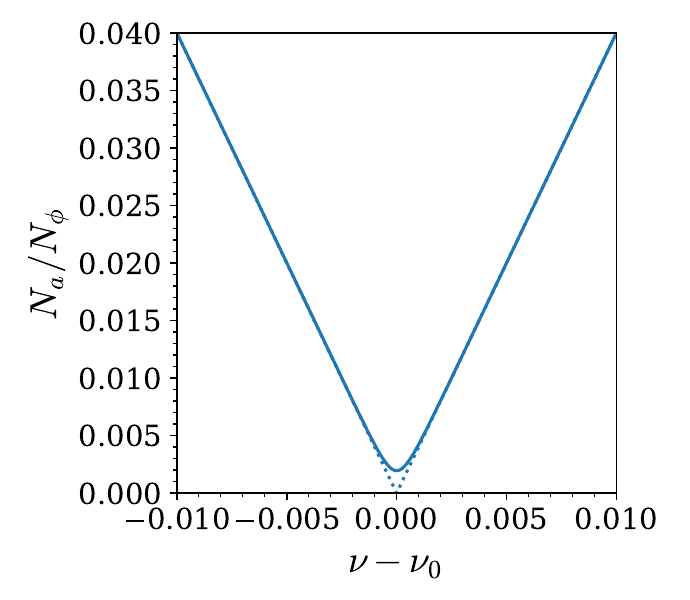}
\caption{
Total number of non-abelian excitations at fillings close to $\nu_0 = -1 + 1/2$. The dashed line shows the expectation $N_a = 4 |\Delta N|$ in the absence of inhomogeneous broadening. We take the broadening parameter $\Gamma$ from fitting the experimental data.}
\label{fig:Na}
\end{figure*}

\end{document}